\documentclass[letterpaper]{JHEP3}
\usepackage{epsfig}

\newcommand{\roughly}[1]{\mathrel{\raise.3ex\hbox{$#1$\kern-0.85em
\lower1ex\hbox{$\sim$}}}}

\def\pd{\partial}
\newcommand{\abs}[1]{\left|{#1}\right|}

\def\cA{{\cal A}}
\def\cB{{\cal B}}
\def\cQ{{\cal Q}}

\def\cF{{\cal F}}
\def\cH{{\cal H}}

\def\cM{{\cal M}}
\def\cO{{\cal O}}
\def\cR{{\cal R}}
\def\cT{{\cal T}}
\def\cU{{\cal U}}

\def\exd{{\hbox{d}}}

\def\ba{\begin{eqnarray}}
\def\ea{\end{eqnarray}}
\def\be{\begin{equation}}
\def\ee{\end{equation}}

\def\ssJ{{\scriptscriptstyle J}}

\def\ssM{{\scriptscriptstyle M}}
\def\ssN{{\scriptscriptstyle N}}
\def\ssP{{\scriptscriptstyle P}}
\def\ssQ{{\scriptscriptstyle Q}}
\def\ssR{{\scriptscriptstyle R}}
\def\ssS{{\scriptscriptstyle S}}

\def\JF{{\scriptscriptstyle J \kern-0.15em F}}
\def\EF{{\scriptscriptstyle E \kern-0.15em F}}

\def\cO{\mathcal{O}}
\def\cH{\mathcal{H}}

\def\cF{\mathcal{F}}
\def\cV{\mathcal{V}}

\def\nn{\nonumber}

\def\({\left(}
\def\){\right)}

\def\pref#1{(\ref{#1})}
\def\eff{{\rm eff}}

\def\pphi{v}
\def\vaceng{\varrho_\eff}

\title{Large Dimensions and Small Curvatures from Supersymmetric Brane Back-reaction}

\author{ C.P.~Burgess${}^{1,2}$ and L.~van Nierop${}^1$\\

${}^1$Department of Physics \& Astronomy, McMaster University\\ \qquad 1280 Main Street West, Hamilton ON, Canada.\\

${}^2$Perimeter Institute for Theoretical Physics\\
\qquad 31 Caroline Street North, Waterloo ON, Canada.\\
}

\date{}

\abstract {We compute the back-reaction of pairs of codimension-two branes within an explicit flux-stabilized compactification, to trace how its properties depend on the parameters that define the brane-bulk couplings. Both brane tension and magnetic couplings to the stabilizing flux play an important role in the resulting dynamics, with the magnetic coupling allowing some of the flux to be localized on the branes (thus changing the flux-quantization conditions). We find that back-reaction lifts the classical flat directions of the bulk supergravity, and we calculate both the scalar potential and changes to the extra-dimensional and on-brane geometries that result, as functions of the assumed brane couplings. When linearized about simple rugby-ball geometries the resulting solutions allow a systematic exploration of the system's response. Several of the systems we explore have remarkable properties. Among these are a propensity for the extra dimensions to stabilize at exponentially large sizes, providing a mechanism for generating extremely large volumes. In some circumstances the brane-dilaton coupling allows the bulk dilaton to adjust to suppress the on-brane curvature parametrically below the change in brane tension, potentially providing a mechanism for reducing the vacuum energy. We explore the stability of this suppression to quantum effects in the case where their strength is controlled by the value of the field along the classical flat direction, and find it can (but need not) be stable. }

\begin{document}

\section{Introduction}

Most of what is known about the physics of branes situated within extra dimensions either neglects their back-reaction onto their environment, or approximates the surrounding geometry as noncompact by ignoring the physics responsible for its stabilization at finite volume.\footnote{Randall-Sundrum models \cite{RS} are important exception to this statement, where back-reaction is incorporated through the Israel junction conditions \cite{IJC}, but these are restricted to the limiting special case of codimension-one branes.} Although these are often good approximations, there are also very interesting situations where they are not.

A particularly interesting case where these effects cannot be neglected is when it is the back-reaction itself that stabilizes some of the extra-dimensional moduli. This case turns out to be important for compactifications whose extra-dimensional volume is very large, such as those arising within large-volume string vacua \cite{LVS}. In particular, the larger the extra dimensions the lower the string scale \cite{LEstrings}, and once the string scale gets as low as the TeV scale --- such as in supersymmetric extensions \cite{Towards} of ADD-type models \cite{ADD} --- supersymmetry becomes dominantly broken on the branes rather than by the fluxes in the bulk \cite{SUSYADD,MSLED}. In this case it is known that brane-induced corrections can dominate the leading classical predictions for the potential governing the lightest moduli \cite{uber}.

The need to include back-reaction when computing the shape of the low-energy scalar potential is both a potential asset and a liability. The downside is the additional complexity required to properly incorporate both the extra-dimensional and brane dynamics within a controlled approximation. The upside is the potential for progress finding new mechanisms for understanding long-standing problems. Progress in particular on naturalness problems to do with the existence of light scalar masses and small vacuum energies, that hinge on understanding {\em all} contributions relevant to the low-energy scalar potential. And there are a variety of reasons for thinking that brane dynamics could be useful for understanding these problems \cite{preSLED,Towards}.

Six dimensional supergravities provide a fruitful place to explore these issues because they are complicated enough to exhibit many of the features of ten- or eleven-dimensional string vacua, yet they are simple enough often to allow explicit solutions and so more systematic exploration of the various configurations of physical interest. 6D gauged chiral supergravity \cite{NS} has proven particularly useful, providing early insights into chiral fermions and flux compactifications \cite{SSs,6Dflux}. This has motivated finding a great many exact solutions to the classical field equations for this system, including a broad class of flux compactifications for which the two extra dimensions are a warped, squashed sphere with singularities at the positions of two positive-tension source 3-branes. These include solutions for which the on-brane geometry is flat \cite{Towards,GGP,GGPplus} (also known as `rugby-ball' solutions), de Sitter/anti de Sitter like \cite{6DdS}, time dependent \cite{TimeDep} or involves other bulk fields \cite{Swirl} or additional branes \cite{MultiBranes}.

In this paper we explore brane back-reaction in this system by computing how the flat rugby-ball solutions respond to a general perturbation of the brane-bulk couplings. In particular, we assume the perturbed brane-bulk couplings to be given by the leading terms in a derivative expansion of the brane action,
\be
\label{braneaction1}
 S_b = \int_{\Sigma_b}  \Bigl( \tau_b \, \omega + \Phi_b \, {}^\star \cF  \Bigr) \,,
\ee
where $\omega$ is the volume form for the space-filling 3-brane, and ${}^\star \cF$ denotes the 6D Hodge dual of the background Maxwell flux, $\cF_{\ssM \ssN}$ (whose presence stabilizes some of what would otherwise be light bulk moduli, in the same way that 3-form fluxes stabilize some moduli in ten-dimensional flux compactifications \cite{GKP}). The coefficient $\tau_b$ denotes the tension of the brane in question, which can be an arbitrary function of the bulk scalar dilaton, $\phi$, appearing in 6D gauged chiral supergravity. $\Phi_b$ has a similar interpretation \cite{Towards,BvN} as an on-brane flux, and can sometimes compete with $\tau_b$ to play an important role in the low-energy energetics of the back-reaction.

The bulk geometry that interpolates between a generic pair of source branes is known to be time dependent \cite{TimeDep}, in much the same way that a random collection of mutually interacting electric charges is also not static. This is reflected by the generic absence of time-independent solutions once a brane-bulk system is perturbed. Unlike earlier stability analyses for these geometries \cite{stability}, we do not deal with this by seeking the time-dependence of the solutions to the brane-perturbed bulk equations of motion. Rather, we instead couple an external current that stabilizes this time-dependence in order to study the energetics of the potential energy that drives it. In practice, at low energies this current need only couple to the massless Kaluza-Klein (KK) `breathing' mode of the leading-order extra-dimensional geometry, since this is a flat direction in field space along which the time dependence dominantly lies.

In this way we find the response of the on- and off-brane geometries as a function of the perturbing brane couplings, as well as the shape of the scalar potential that stabilizes and gives a mass to the low-energy breathing mode, for general choices for the brane coupling functions $\tau_b$ and $\Phi_b$. We find instances where the breathing mode is stabilized by the interaction of the branes on the bulk, as well as cases where it instead runs away to infinity (which, perhaps surprisingly, includes the simplest case where both $\tau_b$ and $\Phi_b$ are independent of the 6D bulk dilaton, $\phi$).

When restricted to the special cases for which our results duplicate earlier calculations, we fully reproduce earlier expressions. But our systematic survey of perturbed solutions also reveals some new ones with surprising properties. These include (see \S4\ for a more detailed summary):
\begin{itemize}
\item Solutions whose extra-dimensional volumes stabilize at values that exponentiate any moderately large hierarchies among the brane-bulk couplings, naturally giving enormously large volumes;
\item Solutions whose on-brane geometry can be parametrically small compared with the largest energy scales that appear in the brane-bulk couplings (though, alas, not yet small enough to describe the observed Dark Energy density);
\item Solutions for which the value of the breathing mode along the low-energy flat direction defines the strength of both brane and bulk loop corrections, and for which this ensures that the above two properties can be stable against quantum effects;
\item Models for which the brane-bulk couplings can have the form required to profit from a `chameleon' mechanism \cite{chameleon}.
\end{itemize}

Our presentation is organized as follows: The next section, \S2, describes the linearized solutions to the bulk field equations, and how the integration constants in these solutions are determined by matching to the functions $\tau_b$ and $\Phi_b$ that define the codimension-2 bulk-brane interactions. These are then used to provide explicit expressions for the extra-dimensional and on-brane geometries as functions of these brane properties. The results of the full 6D calculation are compared with the effective 4D picture that captures the low-energy limit, since the scalar potential in this effective theory provides an efficient way to understand the implications of brane dynamics on low-energy properties. This section closely follows the logic of ref.~\cite{BvN}, which performs a similar calculation in the non-supersymmetric case.

\S3\ then uses the general results of \S2\ to explore the implications of several simple illustrative choices for the coupling functions $\tau_b$ and $\Phi_b$. A particularly simple toy model --- for which $\sum_b \tau_b \propto \sum_b \Phi_b \propto \phi^\eta$, for small $\eta$ --- is also examined, that exhibits modulus stabilization at exponentially large volume and parametric suppression of the low-energy on-brane curvature (or vacuum energy). Finally, this section estimates the effects of brane and bulk loops for the toy model, and argues that the exponentially large volume, and the small on-brane vacuum energy (and scalar masses) can be technically natural.

Our conclusions are summarized in \S4.

\section{The bulk-brane system}

This section defines the system of interest. The fields of interest are part of the bosonic sector of chiral gauged supergravity in six dimensions \cite{NS}, for which we follow the implications of coupling to nonsupersymmetric branes. In particular we follow the metric, $g_{\ssM\ssN}$; a bulk Maxwell gauge potential, $\cA_\ssM$, whose presence helps stabilize the bulk geometry; and the 6D scalar dilaton, $\phi$.

\subsection{Field equations and background solutions}

 We first describe the bulk equations of motion and brane boundary conditions, followed by a simple class of rugby-ball solutions near which general solutions are sought.

\subsubsection*{Bulk equations}

The bosonic action in the bulk is\footnote{We use a `mostly plus' metric and Weinberg's curvature conventions \cite{Wbg} (that differ from those of MTW \cite{MTW} only by an overall sign in the definition of the Riemann tensor).}
\be \label{BulkAction}
 S_\mathrm{bulk} = - \int \exd^6 x \sqrt{-g} \; \left\{ \frac1{2\kappa^2} \, g^{\ssM\ssN}
 \Bigl( \cR_{\ssM \ssN} + \pd_\ssM \phi \, \pd_\ssN \phi \Bigr)
 + \frac14 \, e^{-\phi} \cF_{\ssM\ssN} \cF^{\ssM\ssN}
 + \frac{2 \, g_\ssR^2}{\kappa^4} \, e^\phi \right\} \,,
\ee
where the two dimensionful constants are the gauge coupling, $g_\ssR$, for a specific $U_\ssR(1)$ symmetry of the supersymmetry algebra, and the 6D gravitational constant, $\kappa$. One of these sets the overall scale of the bulk physics, leaving the dimensionless combination $g_\ssR^2/\kappa$ as a free parameter. Here $\cF = \exd \cA$ denotes the gauge potential's field strength.

The equations of motion from this action are the (trace reversed) Einstein equations
\be \label{BulkEinsteinEq}
 \cR_{\ssM\ssN} + \partial_\ssM \phi \, \partial_\ssN \phi
  + \kappa^2  e^{-\phi}\cF_{\ssM \ssP} {\cF_\ssN}^\ssP
  - \left( \frac{\kappa^2}{8} \,
  e^{-\phi} \cF_{\ssP\ssQ} \cF^{\ssP \ssQ}
  - \frac{g_\ssR^2}{\kappa^2} \, e^\phi  \right)
  g_{\ssM\ssN} = 0 \,,
\ee
the Maxwell equation
\be \label{BulkMaxwellEq}
 \nabla_\ssM (e^{-\phi}\cF^{\ssM \ssN}) = 0 \,,
\ee
and the dilation equation
\be \label{BulkDilatonEq}
 \Box \phi - \frac{2 \, g_\ssR^2 }{\kappa^2} \, e^\phi  + \frac{\kappa^2}4 \, e^{-\phi}
 \cF_{\ssM\ssN} \cF^{\ssM\ssN} = 0 \,.
\ee
Since these field equations are invariant under the transformations
\be \label{classscaleinv}
 g_{\ssM \ssN} \to \zeta \, g_{\ssM \ssN} \quad \hbox{and} \quad
 e^{-\phi} \to \zeta \, e^{-\phi} \,,
\ee
with $\cA \to \cA$, any nonsingular solution is always part of a one-parameter family of solutions that are exactly degenerate (within the classical approximation).

\subsubsection*{Symmetry ansatz}

In what follows we restrict attention to solutions that have maximal symmetry in the four on-brane directions, and axial symmetry in the two extra dimensions. This assumption restricts us to solutions involving at most two source branes. The corresponding \emph{ans\"atze} for the metric and Maxwell field are
\be
\label{bulk-brane system: ansatz}
 \exd s^2 = \exd \rho^2 + e^{2B} \exd\theta^2 + e^{2W} \,
 \hat g_{\mu\nu} \exd x^\mu \exd x^\nu
 \quad \hbox{and} \quad
 \cA = \cA_\theta \, \exd \theta \,,
\ee
where $\hat g_{\mu\nu}(x)$ is a maximally symmetric metric, and all of the functions $W$, $B$, $\phi$ and $\cA_\theta$ depending only on $\rho$. The corresponding Maxwell field strength is $\cF_{\rho\theta} = \cA_\theta'$, where primes denote differentiation with respect to the coordinate $\rho$.

Subject to this \emph{ansatz} the bulk field equations reduce to
\ba \label{eomansatz}
 \left( e^{-B + 4W} e^{-\phi} \cA_{\theta}' \right)' &=& 0
 \qquad(\cA_\theta) \label{Ade} \nn\\
 \left( e^{B+4W} \phi' \right)' - \left( \frac{2g_\ssR^2}{\kappa^2} \, e^\phi
 - \frac12\, \kappa^2 \cQ^2 \, e^\phi e^{-8W} \right) e^{B+4W} &=& 0 \qquad(\phi) \nn\\
 4 \Bigl[ W''+(W')^2 \Bigr] + B'' + (B')^2 + (\phi')^2 + \frac34 \, \kappa^2 \cQ^2
 \, e^\phi e^{-8W} + \frac{g_\ssR^2}{\kappa^2} \, e^\phi &=& 0 \qquad(\rho\rho) \\
 B'' + (B')^2 + 4W'B' + \frac34 \, \kappa^2 \cQ^2 \, e^\phi e^{-8W}
 + \frac{g_\ssR^2}{\kappa^2} \, e^\phi &=& 0 \qquad(\theta\theta) \nn\\
 \frac14 \, e^{-2W} \hat R + W'' + 4(W')^2 + W'B' - \frac14 \, \kappa^2\cQ^2 \, e^\phi e^{-8W}
 +\frac{g_\ssR^2}{\kappa^2} \, e^\phi &=& 0 \qquad(\mu\nu) \,. \nn
\ea
The first of these can be integrated once exactly, introducing an integration constant, $\cQ$, labeling the bulk flux,
\be \label{Afirstintegral}
 \cF_{\rho\theta}  = \cA_\theta'
  = \cQ \, e^\phi e^{B-4W}  \,.
\ee

\subsubsection*{Rugby ball solutions}

In the special case that the dilaton is constant, $\phi = \varphi_0$, these equations have a simple solution with extra dimensions having the shape of a rugby ball, sourced by two branes \cite{Towards}:
\ba \label{symmansatz}
 \exd s^2 &=& e^{-\varphi_0} \left[ \exd \hat \rho^2
 + \alpha^2 L^2 \sin^2 \left( \frac{\hat\rho} L \right)
 \exd\theta^2 \right] + \eta_{\mu\nu} \exd x^\mu \exd x^\nu \nn\\
 \cF_{\rho\theta} &=& \cF_{\hat\rho\theta} \, e^{-\varphi_0/2}
 = \cQ e^{\varphi_0/2} \alpha L \sin \left( \frac{\hat\rho} L \right) \,,
\ea
where $\hat g_{\mu\nu} = \eta_{\mu\nu}$ denotes the usual flat metric of Minkowski space. The extra-dimensional metric becomes singular at the brane positions, $\hat\rho_\ssN = 0$ and $\hat\rho_\ssS = \pi L$, which are $\varphi_0$-independent because of the coordinate rescaling $\rho := e^{-\varphi_0/2} \hat\rho$.

The geometry generically has a conical singularity at these points, characterized by the defect angle $\delta = 2\pi(1 - \alpha)$. In the special case $\alpha = 1$ the extra-dimensional geometry is a sphere, corresponding to the supersymmetric Salam-Sezgin solution \cite{SSs}. The deficit angle can be related to the common tension, $T$, of the two source branes by \cite{TvsA}
\be
 1 - \alpha = \frac{\kappa^2 T}{2\pi} \,.
\ee

The equations of motion impose two relations amongst the integration constants, requiring
\ba \label{rugbyrelns}
 \frac{2 g_\ssR^2}{\kappa^2} &=& \frac{\kappa^2\cQ^2}{2} \quad \hbox{(dilaton equation)} \nn\\
 \hbox{and} \quad
 \kappa^2 \cQ^2 L^2 &=& 1 \qquad\quad \hbox{(Einstein equation)} \,.
\ea
Additionally, flux quantization due to the spherical topology of the extra dimensions implies
\be \label{rugbyfluxquant}
 \frac ng = 2\alpha L^2\cQ = \frac\alpha{g_\ssR} \,,
\ee
where $g$ is the gauge coupling of the background Maxwell field and $n$ is an integer. The couplings $g$ and $g_\ssR$ are in general different because the background Maxwell field need not be the one that gauges the $U_\ssR(1)$ symmetry. This last condition determines the deficit angle, $\alpha$, and thereby constrains the tension of the source branes. As is elaborated in more detail below, a minor modification \cite{Towards} of these solutions allows the source branes themselves to carry some of the total flux, $\Phi_{\rm branes}$, in which case eq.~\pref{rugbyfluxquant} generalizes to
\be \label{rugbyfluxquantwbranes}
  \frac{n}{g} = \frac{\alpha}{g_\ssR} + \frac{\Phi_{\rm branes}}{2\pi}
   = \frac{1}{g_\ssR}\left( 1 - \frac{\kappa^2 T}{2\pi} \right)
    + \frac{\Phi_{\rm branes}}{2\pi}
        = \frac{1}{g_\ssR}\left[ 1 - \frac{\kappa^2 }{2\pi} \Bigl(
    T - \frac{\cQ \, \Phi_{branes}}{2} \Bigr) \right] \,,
\ee
where the last equality uses eqs.~\pref{rugbyrelns}, which imply $\cQ = 2 g_\ssR/\kappa^2$. This can be regarded as allowing the tension in these solutions to be arbitrary, provided the on-brane flux is also dialed, $\Phi_{\rm branes}(T)$, to satisfy eq.~\pref{rugbyfluxquantwbranes}.

For fixed brane flux the above construction describes only a one-parameter family of solutions, labeled by $\varphi_0$. This one-parameter degeneracy is the one required by the scale invariance, eq.~\pref{classscaleinv}, of the classical field equations. Because of the overall factor of $e^{-\varphi_0}$ in the extra-dimensional metric, eq.~\pref{symmansatz}, the proper distance between the two branes is $\Delta \rho = e^{-\varphi_0/2} \pi L$ and the volume of the extra dimensions is
\be \label{XDvolume}
 \cV_2 = 4 \pi \alpha L^2 e^{-\varphi_0} \,.
\ee

Our interest in what follows is in how this flat direction gets lifted by dilaton couplings to the branes. Its connection to the extra-dimensional volume makes this also a stabilization mechanism for the size of the extra dimensions; a codimension-2 generalization of the better-known Goldberger-Wise stabilization mechanism for codimension-1 branes \cite{GW} within RS models.

\subsubsection*{Brane matching conditions}

We take the brane-bulk coupling to be defined by the following lowest-derivative action, including both a $\phi$-dependent tension and a $\phi$-dependent coupling to the Maxwell field \cite{Towards}:
\ba \label{BraneFluxCoupling}
 S_\mathrm{branes} &=& - \sum_{b=\ssN,\ssS} \int d^4x \sqrt{-g_4} \;
 \left[ \tau_b - \frac12 \, \Phi_b \, e^{-\phi} \epsilon^{mn} \cF_{mn} \right] \nn\\
 &=& - \sum_{b=\ssN,\ssS} \int d^4x \sqrt{-\hat g} \; e^{4W}
 \left[ \tau_b - \frac12 \, \Phi_b \, e^{-\phi} \epsilon^{mn} \cF_{mn} \right] \,,
\ea
where the coupling functions $\tau_b$ and $\Phi_b$ can depend on all of $\phi$, $W$ and $g_{\theta\theta}$ without breaking the condition of maximal symmetry in the on-brane directions. Because of the explicit factor of $e^{-\phi}$ extracted from the Maxwell coupling, these interactions also do not break the bulk scaling symmetry, eq.~\pref{classscaleinv}, only when both $\tau_b$ and $\Phi_b$ are $\phi$-independent. Our conventions are such that $\epsilon^{\rho \theta} = 1/\sqrt{g_2} = e^{-B}$ transforms as a tensor, rather than a tensor density, in the two transverse dimensions. The parameter $\tau_b$ has the physical interpretation of being the tension of the brane, and (as is shown below) the parameter $\Phi_b$ similarly denotes the magnetic charge (or flux) carried by the source branes.

The presence of such brane couplings imposes a set of boundary conditions on the derivatives of the bulk fields in the near-brane limits,\footnote{These matching conditions can be derived \cite{uvcaps} from codimension-1 microscopic models \cite{otheruvcaps} for codimension-2 branes.} given by\footnote{Notice that we normalize the quantities $\cT_b$ and $\cU_b$ without including the factor of $e^{4W}$ used in this reference.} \cite{BBvN}:
\begin{eqnarray} \label{matching}
 \Bigl[ e^{B} {\phi}' \Bigr]_{\rho_b}
 &=& \frac{\partial
 \cT_b}{\partial \phi} \quad \hbox{with} \quad
 \cT_b := \frac{\kappa^2 T_b}{2\pi}  \nn\\
 \Bigl[ e^{B} W' \Bigr]_{\rho_b}
 &=&  \cU_b \quad \hbox{with} \quad
 \cU_b :=  \frac{\kappa^2}{4\pi} \left(
 \frac{\partial T_b}{\partial g_{\theta\theta}} \right)   \\
 \hbox{and } \quad
 \Bigl[ e^B B'-1 \Bigr]_{\rho_b}
 &=& - \Bigl[ \cT_b +  3\,\cU_b
 \Bigr] \,,\nn
\end{eqnarray}
where, as before, primes denote differentiation with respect to $\rho$. $T_b$ is defined in terms of $\tau_b$ and $\Phi_b$ as the total lagrangian density of the source,
\be
 T_b := \tau_b - \, \Phi_b \, e^{-\phi} e^{-B} \cF_{\rho\theta} \,.
\ee

As shown in Appendix \ref{App:fluxconditionws}, the corresponding boundary condition for the Maxwell field implies that the integral of $\cF_{\rho\theta}$ to obtain $\cA_\theta(\rho)$ in a coordinate patch containing each source brane gives
\ba
 \cA_\theta(\rho) &=& \frac{\Phi_\ssN}{2\pi} + \cQ \int_{\rho_\ssN}^{\rho}
 \exd \rho  \; e^{\phi+B-4W} \quad\;\; \hbox{Northern hemisphere} \nn\\
 &=& - \frac{\Phi_\ssS}{2\pi} + \cQ \int_{\rho_\ssS}^{\rho} \exd \rho
 \; e^{\phi+B-4W}  \quad \hbox{Southern hemisphere} \,,
\ea
where $\Phi_b := \lim_{\rho \to \rho_b} \Phi_b[\phi(\rho)]$ --- appropriately renormalized \cite{Bren} --- and the signs are dictated by the observation that increasing $\rho$ points away from (towards) the North (South) pole, together with the requirement that the two patches share the same orientation. Requiring these to differ by a gauge transformation, $g^{-1} \partial_\theta \Omega$, on regions of overlap implies the flux-quantization condition
\be \label{fluxquantzn}
 \frac{n}{g} = \frac{\Phi_{\rm tot}}{2\pi}
 + \cQ \int_{\rho_\ssN}^{\rho_\ssS} \exd \rho \; e^{\phi+B - 4W} \,,
\ee
which identifies $\Phi_{\rm tot} = \sum_b \Phi_b$ as the part of the total magnetic flux carried by the branes \cite{Towards}.

\subsection{Perturbations}

In this section we use the previous discussion to analyze how couplings to the brane lift the flat direction associated with the scaling symmetry of the bulk theory, and so to see how the scalar zero mode, $\varphi_0$, becomes stabilized at a specific value, $\varphi_0 = \varphi_\star$. Our discussion closely follows the discussion of the nonsupersymmetric system in ref.~\cite{BvN}.

It is instructive to contrast how this stabilization differs from the nonsupersymmetric system. To this end recall how the stabilization occurs in detail, from the point of view of six dimensions. Given two branes, we seek the bulk configuration satisfying the field equations that interpolates between the boundary conditions that each brane specifies. Specializing to solutions that are both axially symmetric in the transverse directions and maximally symmetric in the on-brane dimensions requires seeking bulk profiles that depend only on $\rho$.

What is important is that the brane boundary conditions only specify the derivatives of the fields near the branes, and not the values of the fields themselves there. Once the derivatives of the fields are specified at one brane, the values of the fields at the same brane can be adjusted to try to ensure that the derivatives take the values required by the other brane at the other brane's position. It is in this way that the stabilized value, $\varphi_0 = \varphi_\star$, is obtained if the brane actions break the classical bulk scaling symmetry.

This argument shows that a classical solution satisfying all of the boundary conditions is in general impossible given an arbitrary choice for $\varphi_0$. From the low-energy 4D perspective the absence of a solution when $\varphi_0 \ne \varphi_\star$ corresponds to the absence of a static solution for a value of $\varphi_0$ that is not an extremal of the low-energy effective potential, $V_\eff'(\varphi_0) \ne 0$. It can still be possible to map out the shape of the scalar potential for generic $\varphi_0$, however, provided we turn on an external current, $J$, coupled to $\varphi_0$ that is designed to ensure that $\varphi_0$ is a stationary point of the potential, including the current. The shape of the effective potential can be computed by seeing precisely how much current is required as a function of $\varphi_0$. In what follows we define the current coupling by adding the following term to the action\footnote{As is shown in Appendix \ref{App:alternative current}, most of the low-energy physics of interest is insensitive to the detailed form of the current to which we couple, so long as it has a good overlap with the would-be zero mode.}
\be
 S_\ssJ = - \int \exd^6x \sqrt{-g} \; J \,,
\ee
where $J$ is a constant (since our goal is only to couple a current to the would-be zero mode, $\varphi_0$).

In this kind of construction the stabilized value, $\varphi_\star$, corresponds to the choice for which no external current is necessary, $J(\varphi_\star) = 0$. An important difference between the supersymmetric system of interest here and the nonsupersymmetric one studied in ref.~\cite{BvN} is that in the supersymmetric case it can (but need not) happen that there is no value of $\varphi_0$ for which $J(\varphi_0) = 0$. As we shall see, from the 4D point of view this corresponds to an effective potential that is a pure runaway, for which $V_\eff'(\varphi_0)$ only vanishes as $\varphi_0 \to \pm \infty$.

\subsubsection*{Linearized equations}

Our goal is to solve the above field equations by linearizing them about a rugby-ball solution. This amounts to assuming that the $\phi$-dependent contribution to $\tau_b$ is small relative to the tension that is responsible for the rugby-ball geometry itself:
\be
 \tau_b = \tau + \delta \tau_b(\phi)
 \quad \hbox{and} \quad
 \Phi_b = \Phi + \delta \Phi_b(\phi) \,,
\ee
with the background deficit angle sourced by $T = \tau - \cQ \, \Phi$. The linearized equations of motion including the current term --- derived in Appendix~\pref{app:linearization} --- are given below. All background (rugby-ball) quantities are denoted by a subscript 0, and perturbations are universally denoted by $\delta$: so $\cQ = \cQ_0 + \delta\cQ$ {\em etc}. Since we ignore all second-order quantities we may write $\delta \cQ /\cQ \simeq \delta \cQ /\cQ_0$ and so can use either of these quantities interchangeably. Also, since $W_0 = 0$ for the rugby balls, $W = \delta W$.

To linear order the Maxwell field strength becomes
\be \label{linflux}
 \cF_{\rho\theta} = \cQ \alpha L e^{\varphi_0/2}
 \sin \left( \frac{\hat\rho} L \right) \left( 1 +
 \frac{\delta\cQ}{\cQ_0} + \delta B - 4 \delta W \right) \,,
\ee
the on-brane curvature is
\be \label{lincurv}
 \hat R = - 4 e^{\varphi_0} \left[ \frac{2\delta W}{L^2} + \frac1L \cot \left( \frac{\hat\rho} L \right) \partial_{\hat\rho}\, \delta W
 + \partial_{\hat\rho}^2 \, \delta W \right] + \frac{2e^{\varphi_0}}{L^2}
 \left( \frac{\delta\cQ} \cQ \right)
 - 2\kappa^2J \,,
\ee
and the remaining linearized field equations become
\ba \label{linphiBW}
 \partial_{\hat\rho} \left[ \sin\left( \frac{\hat \rho} L \right)
 \partial_{\hat \rho} (\delta \phi) \right]
 &=& \frac{1}{L^2} \left( 4\delta W - \frac{\delta\cQ}\cQ \right)
 \sin \left( \frac{\hat \rho}L \right)\nn\\
 \frac{\partial_{\hat\rho} \left[ \sin^2
 \left( \frac{\hat\rho} L \right) \partial_{\hat\rho}
 (\delta B) \right]}{\sin^2 \left( \frac{\hat \rho}L \right)}
 &=& - \frac{1}{L^2} \left[ \delta\phi
 + \frac32 \left( \frac{\delta\cQ}\cQ \right)
 - 6 \, \delta W + \kappa^2 J L^2 e^{-\varphi_0} \right]
 - \frac4L \, \cot \left( \frac{\hat \rho} L \right)
 \partial_{\hat\rho} \, \delta W \nn\\
 \hbox{and} \quad
 \partial_{\hat\rho}^2 \delta W &=&
 \frac{1}{L} \, \cot \left( \frac{\hat \rho} L \right)
 \partial_{\hat\rho} \, \delta W \,.
\ea

Finally, the linearized flux quantization condition can be expressed as
\be \label{fluxcondition}
 \frac{\delta\cQ}\cQ = \frac{1}{2L} \int_0^{\pi L}
 \exd\hat \rho \; \sin \left( \frac{\hat \rho} L \right)
 \left( 4\delta W - \delta B - \delta\phi \right)
 - \frac{\kappa^2\cQ }{4\pi\alpha}
 \Bigl( \delta \Phi_\ssN  + \delta \Phi_\ssS \Bigr) \,.
\ee

\subsection{Linearized solutions}
The strategy is to construct the general solution to these linearized equations, and then to use the brane matching conditions to eliminate the resulting integration constants in terms of brane properties. To simplify expressions it is convenient to define the dimensionless coordinate $x := \hat\rho/L = (\rho/L) e^{-\varphi_0/2}$, keeping in mind that its implicit dependence on $\varphi_0$ brings this dependence to any bulk fields that depend on $x$. We have some freedom in how to group the perturbations; which we employ (without loss of generality) to simplify the linearized flux-quantization condition as much as possible.\

First, we solve the equation for the warp factor, $W$, which has the general solution
\be
 \delta W(x) = W_0 + W_1 \cos x \,,
\ee
where $W_0$ and $W_1$ are integration constants, of which $W_0 = 0$ may be ensured by rescaling the on-brane coordinates, $x^\mu$.

With this solution, the equation to be solved for the dilaton becomes
\be
 \pd_x \Bigl[ \sin x \, \pd_x ( \delta\phi ) \Bigr] =
 \left( 4 W_1 \cos x - \frac{\delta\cQ}\cQ \right) \sin x \,,
\ee
which integrates to give
\be \label{linearphisoln}
 \delta \phi\,(x) = \delta \varphi_0 + \varphi_1
 \ln \left| \frac{1-\cos x}{\sin x} \right| - 2W_1 \cos x + \left(
 \frac{\delta\cQ}{\cQ} \right) \ln | \sin x | \,.
\ee
Here $\delta \varphi_0$ and $\varphi_1$ are integration constants, of which $\delta \varphi_0 = 0$ can be ensured without loss of generality by absorbing it into the otherwise arbitrary background value, $\varphi_0$.

Finally, the equation of motion for $\delta B$ becomes
\ba
 \frac{\pd_x \left[ \sin^2 x \, \pd_x (\delta B ) \right] }{ \sin^2 x} &=&
 - \varphi_1 \ln \left| \frac{1-\cos x }{\sin x} \right|
 - \frac{\delta\cQ}\cQ \left( \frac32 + \ln | \sin x | \right) \\
 &&\qquad\qquad\qquad\qquad
 + 12 W_1 \cos x - \kappa^2 J L^2 e^{-\varphi_0} \,.\nn
\ea
This integrates to
\ba
 \delta B &=& \delta \hat B_0 + B_1 \cot x -4 W_1 \cos x
 -  \varphi_1 \cM_2(x) \\
 && \qquad\qquad\qquad
 + \frac{\delta\cQ}\cQ \left[ \frac{3}{4} \,x \cot x - \cH_2(x) \right]
 + \frac12 \left( \kappa^2 J L^2 e^{-\varphi_0}
 \right)  x \cot x \,, \nn
\ea
where $\delta \hat B_0$ and $B_1$ are integration constants. Of these, $B_1$ is pure gauge in that it can be changed arbitrarily by reparameterizing the coordinate $\rho$. We fix this coordinate freedom by defining $\rho = 0$ to be the position of the `north' brane, which requires $e^B \to 0$ as $\rho \to 0$; ensuring $B_1 = 0$. The functions $\cM_2$ and $\cH_2$ appearing here are defined by
\ba
 \cM_1(x) &:=& \int_0^x \exd y \, \sin^2 y \ln \left| \frac{1-\cos y}{\sin y}\right| \nn\\
 \cM_2(x) &:=& \int_0^x \exd y \; \frac{\cM_1(y)}{\sin^2 y}   \,,
\ea
and
\ba
 \cH_1(x) &:=& \int_0^x \exd y \, \sin^2 y \, \ln |\sin y | \nn\\
 \cH_2(x) &:=& \int_0^x \exd y \; \frac{\cH_1(y)}{\sin^2 y} \,.
\ea

For later convenience when discussing flux quantization it is useful to absorb parts of these integrals into the definition of $\delta \hat B_0$, by writing
\ba \label{newdeltaB0}
 \delta B &=& \delta B_0 -4 W_1 \cos x + \varphi_1 \left[ \frac{\overline\cM}2
 - \cM_2(x) \right] + \frac{\delta\cQ}\cQ \left[ \frac{3}{4} \, x\cot x
 - \cH_2(x) \right] \nn\\
 && \qquad\qquad + \frac12 \left( \kappa^2 J L^2 e^{-\varphi_0} \right)
 (x \cot x +1) \,,
\ea
with the number $\overline\cM$ defined by
\be
 \overline\cM := \int_0^\pi \exd x \, \sin x \, \cM_2(x) \,.
\ee
Numerically this evaluates to the value\footnote{Maple 11, 10 digit precision, see Appendix \pref{app:findM}} $\overline \cM = -1$, which we use throughout what follows.

\subsubsection*{Flux quantization}

Using the above expressions in the linearized flux quantization condition, eq. ~\pref{fluxcondition}, gives
\ba \label{fluxrepeat}
 \frac{\delta\cQ}\cQ&=&\frac12\int_0^\pi \exd x \, \sin x \Bigl(
 4 \delta W - \delta B - \delta \phi \Bigr) - \frac{\kappa^2\cQ}{4\pi\alpha}
 \Bigl( \delta \Phi_\ssN + \delta \Phi_\ssS \Bigr) \nn\\
 &=& - \delta B_0 + \frac34 \left( \frac{\delta\cQ}\cQ \right)
 - \frac{\kappa^2\cQ}{4\pi\alpha} \Bigl( \delta \Phi_\ssN + \delta \Phi_\ssS \Bigr) \,,
\ea
which uses the integral
\be
 \overline \cH := \int_0^\pi \exd x \, \sin x \, \cH_2(x)
 \simeq -0.613706 \simeq \ln 4 - 2 \,,
\ee
and the last approximate equality is a numerical inference.\footnote{Mathematica 7, with thanks to Ben Jackel.} The absence of $\varphi_1$ on the right-hand side of eq.~\pref{fluxrepeat} is a consequence of the definition of $\delta B_0$ used in eq.~\pref{newdeltaB0}. Solving this for $\delta B_0$ gives
\be \label{dB0soln}
  \delta B_0  = - \frac14 \left( \frac{\delta\cQ}\cQ \right)
  - \frac{\kappa^2\cQ}{4\pi\alpha}
 \Bigl( \delta \Phi_\ssN + \delta \Phi_\ssS \Bigr) \,.
\ee

Finally, the linearized field equations return the following on-brane curvature,
\ba
 \hat R &=& -4 e^{\varphi_0} \left[ \frac{2\delta W}{L^2} + \frac1L \, \cot \left(
 \frac{\hat\rho} L \right) \partial_{\hat\rho}\, \delta  W
 + \partial_{\hat\rho}^2 \, \delta W \right] + \frac{2e^{\varphi_0}}{L^2} \left(
 \frac{\delta\cQ}\cQ \right) - 2\kappa^2 J  \nn\\
 &=& \frac{2 e^{\varphi_0}}{L^2} \left( \frac{\delta\cQ}\cQ \right) - 2\kappa^2 J \,.
\ea
Notice that all of the $\rho$-dependence cancels in this expression (as must happen given our assumption of maximal symmetry), leaving a result that is determined purely by the change of bulk Maxwell flux and the applied current.

\subsection{Physical interpretation and renormalization}

The above solutions are described by four physical integration constants, which we can take to be $\varphi_0$, $\varphi_1$, $W_1$ and $\delta \cQ/\cQ$. These can be traded for four physical properties of the bulk and on-brane geometries.

$W_1$ can be taken to be the difference between the value of the warping (which controls the gravitational redshift) between the two branes, which is given by
\be
 \delta W_\ssN - \delta W_\ssS = 2 \, W_1  \,.
\ee

Similarly, to linear order the near-brane geometry as $x = \hat \rho/L \to 0$ is governed by
\ba
 e^B &\simeq& e^{-\varphi_0/2} \alpha L \sin x \Bigl[ 1 + \delta B(x) \Bigr] \nn\\
 &\simeq& \alpha \rho \left[ 1 + \delta B_0 - 4W_1 - \frac{\varphi_1}{2} + \frac34 \left( \frac{\delta \cQ}\cQ \right) + \kappa^2 J L^2 e^{-\varphi_0}
 + \cO(\rho^2) \right]\,,
\ea
which corresponds to a conical singularity (since the $e^B$ vanishes linearly with $\rho$), having defect angle $\alpha_\ssN = \alpha + \delta\, \alpha_\ssN$ with
\be
 \frac{\delta \,\alpha_\ssN} \alpha = \delta B_0 - 4 W_1 - \frac{\varphi_1} 2
 + \frac34 \left( \frac{\delta\cQ}\cQ \right) + \kappa^2 J L^2 e^{-\varphi_0} \,.
\ee

By contrast, as $x = \hat \rho/L \to \pi$ we have
\be \label{eBlimpi}
 e^B \to \pi\alpha Le^{-\varphi_0/2} \left[ \left(1 - \frac12 \, \ln 2 \right)
 \left( \frac{\delta\cQ}\cQ \right)
 - \frac12 \left( \kappa^2 J L^2 e^{-\varphi_0} \right) \right] \,,
\ee
which uses $\cH_2(\pi - \varepsilon) = (1 - \ln 4) ({\pi}/{4 \varepsilon}) + \cO(\varepsilon^0)$. In particular, this shows that $e^B$ does {\em not} vanish at $\hat \rho = \pi L$. Instead $e^B$ vanishes at $\hat \rho = \pi (L + \delta L)$, indicating a change in proper distance between the branes: $\rho_\ssS - \rho_\ssN = \pi (L + \delta L) e^{-\varphi_0/2}$. The amount of the change is obtained by comparing eq.~\pref{eBlimpi} to the Taylor expansion of $e^B$ about its new zero, giving
\be
 \frac{\delta L}L \simeq - \left[ \frac34 \left( \frac{\delta\cQ}\cQ \right) + \frac12 \left( \kappa^2 J L^2 e^{-\varphi_0} \right) \right]
 \,.
\ee

The singularity at the `south' brane is also conical (at linear order), with defect angle given by $\alpha_\ssS = \alpha + \delta \, \alpha_\ssS$ with
\be
 \frac{\delta \,\alpha_\ssS} \alpha =  \delta B_0 + 4 W_1 + \frac{\varphi_1} 2
 + \frac34 \left( \frac{\delta\cQ}\cQ \right) + \kappa^2 J L^2 e^{-\varphi_0} \,.
\ee

One might imagine that a further observable could be the difference between the value of the dilaton field at the two branes, $\phi_\ssN - \phi_\ssS$, since this governs the relative strength of some of the bulk couplings to each brane (such as the strength of the bulk Maxwell couplings to brane-localized charged particles). However a subtlety arises in this case because the profile $\phi(\rho)$ diverges in the limit that $\rho \to \rho_\ssN$ and $\rho \to \rho_\ssS$. For this reason we defer a discussion of this quantity to the next section, which deals with renormalizing these divergences.

\subsubsection*{Brane matching and renormalization}\label{renormalization}

Ultimately the bulk integration constants should be related to physical properties of the branes that are the source of the bulk geometry; this is where the brane matching conditions play a role. In order to perform this matching we must specify a functional form for the brane tensions, $\tau_b$, and fluxes, $\Phi_b$. We take both of these to be smooth functions of $\phi$, and in many (but not all) examples we imagine these functions to be extremized at $\phi = \hat \pphi_b$: that is, $(\partial \tau_b/\partial \phi)_{\phi = \hat \pphi_b} = 0$.

The problem in practice with matching is that the argument of $\tau_b$ and $\Phi_b$ is $\phi_b := \phi(\rho_b)$, but the profile $\phi(\rho)$ given in eq.~\pref{linearphisoln} diverges as $\rho \to \rho_b$. For instance, for $x = \hat \rho / L = \varepsilon \ll 1$ and $x = \pi - \varepsilon$ we have
\ba
 \phi\,(x = \varepsilon) &=& \varphi_0  - 2W_1 + \varphi_1
 \ln \left| \frac{\varepsilon}{2} \right| + \left(
 \frac{\delta\cQ}{\cQ} \right) \ln | \varepsilon | + \cO(\varepsilon)\nn\\
 \hbox{and} \quad
 \phi\,(x = \pi - \varepsilon) &=& \varphi_0  + 2W_1 + \varphi_1
 \ln \left| \frac{2}{\varepsilon} \right| + \left(
 \frac{\delta\cQ}{\cQ} \right) \ln | \varepsilon |  + \cO(\varepsilon) \,.
\ea

This divergence is dealt with by renormalizing the parameters that define the functional form of $\tau_b$ and $\Phi_b$, and in particular those parameters that determine the values $\hat\pphi_b$. It can be absorbed in the definitions of $\hat\pphi_b$ by defining renormalized quantities, $\pphi_b$:
\ba
 \pphi_\ssN &=& \hat\pphi_\ssN - \frac{\delta\cQ}\cQ \,
 \ln \left( \varepsilon \right) - \varphi_1 \ln \left( \varepsilon/2 \right) \nn\\
 \pphi_\ssS &=& \hat\pphi_\ssS - \frac{\delta\cQ}\cQ\,
 \ln \left( {\varepsilon} \right) + \varphi_1
 \ln \left( {\varepsilon/2} \right) \,,
\ea
where the first expression is relevant at $x=0$ (the north brane positon), and the second one at $x=\pi$ (the south brane). With these definitions,
\ba
 \lim_{\varepsilon \to 0} \Bigl[ \phi(\varepsilon) - \hat\pphi_\ssN
 \Bigr] &=& \varphi_0 - 2W_1 - \pphi_\ssN \nn\\
 \hbox{and} \quad
 \lim_{\varepsilon \to 0} \Bigl[ \phi( \pi - \varepsilon) - \hat\pphi_\ssS
 \Bigr] &=& \varphi_0 + 2W_1 - \pphi_\ssS \,,
\ea
and so $\tau_b(\phi - \hat \pphi_b) = \tau_b(\varphi_0 \pm 2W_1 - \pphi_b)$ and so on. This is a useful redefinition because our interest really is in the value at which the zero mode, $\varphi_0$, gets stabilized, rather than on the value of $\phi$ itself at the brane position. And this is finite as $\varepsilon \to 0$ with renormalized quantities (like $\pphi_b$) fixed.

With this construction in mind, there are four independent matching conditions:
\ba
 \left[ e^B \partial_\rho \phi \right]_{\rho = 0} = \frac{\kappa^2 }{2\pi}
 \left( \frac{\partial T_\ssN}{\partial \phi} \right)
 \quad &\hbox{and}& \quad
 \left[ e^B \partial_\rho \phi \right]_{\rho=\pi L} = -\frac{\kappa^2 }{2\pi}
  \left( \frac{\partial T_\ssS}{\partial \phi} \right) \nn\\
 \left[ e^B  \partial_\rho B \right]_{\rho=0} = 1 - \frac{\kappa^2 T_\ssN}{2\pi}
 \quad &\hbox{and}& \quad
 \left[ e^B  \partial_\rho B \right]_{\rho=\pi L} =
 -1 + \frac{\kappa^2 T_\ssS}{2\pi} \,,
\ea
where, as before, $T_b = \tau_b - \cQ \, \Phi_b \, e^{-4W(\rho_b)}$. The difference in signs between north and south brane arises because increasing $\rho$ points away from the north brane but towards the south brane.

Specialized to the dilaton profile, eq.~\pref{linearphisoln}, the first two of the above conditions become
\ba
 \alpha \left( \varphi_1 + \frac{\delta\cQ}\cQ \right) &=& \frac{\kappa^2}{2\pi} \left( \frac{\partial \, \delta T_\ssN}{\partial \phi} \right) \nn\\
 \alpha \left( \varphi_1 - \frac{\delta\cQ}\cQ \right) &=& - \frac{\kappa^2}{2\pi}
 \left( \frac{\partial \, \delta T_\ssS}{\partial \phi} \right) \,,
\ea
while the latter two evaluate to
\ba
 \alpha \left[ -4W_1 - \frac{\varphi_1}{2} + \frac34 \left(
 \frac{\delta\cQ}\cQ \right) + \delta B_0 + \kappa^2 J L^2 e^{-\varphi_0} \right]
 &=& -\frac{\kappa^2}{2\pi} \; \delta T_\ssN \nn\\
 \alpha \left[ 4W_1 -\frac{ \varphi_1}{2} +  \frac34 \left(
 \frac{\delta\cQ}\cQ \right) +\delta B_0 + \kappa^2 J L^2 e^{-\varphi_0} \right]
 &=& - \frac{\kappa^2}{2\pi} \; \delta T_\ssS \,,
\ea
where the change in brane action from the background value, $T$, is $\delta T_b := T_b - T = \delta \tau_b - \cQ\, \delta\Phi_b + \cQ \, \Phi [ 4 \delta W (\rho_b) - \delta\cQ/\cQ]$. However, the terms involving $\delta W$ and $\delta \cQ/\cQ$ in $\kappa^2 \delta T_b$ may be dropped in the matching conditions because their contributions are suppressed by an additional factor of $\kappa^2 \cQ \Phi/2\pi$ relative to the leading contributions. Hence, from here on we take $\delta T_b \simeq \delta \tau_b - \cQ\, \delta\Phi_b$.

Eliminating $\delta B_0$ using eq.~\pref{dB0soln}, and solving the above matching conditions gives
\ba
 \frac{\delta\cQ}{\cQ} &=& \frac{\kappa^2}{4\pi\alpha}
 \Bigl[ \delta T_\ssN' + \delta T_\ssS' \Bigr] \nn\\
 \varphi_1 &=& \frac{\kappa^2}{4\pi\alpha}
 \Bigl[\delta T_\ssN' - \delta T_\ssS' \Bigr] \nn\\
 W_1 &=& \frac{\kappa^2}{16\pi\alpha} \left[ \left( \delta T_\ssN + \frac12
 \, \delta T_\ssN' \right) - \left( \delta T_\ssS
 + \frac12 \, \delta T_\ssS' \right) \right] \nn\\
 \kappa^2 J L^2 e^{-\varphi_0} &=& -\frac{\kappa^2}{4\pi\alpha}
 \left[ \left( \delta T_\ssN + \frac12 \, \delta T_\ssN'
 - \cQ \, \delta \Phi_\ssN \right)
 + \left( \delta T_\ssS  + \frac12 \, \delta T_\ssS'
  - \cQ \, \delta \Phi_\ssS \right)\right] \,,
\ea
where $\delta T_b'$ denotes $\partial \,\delta T_b/\partial \phi$. These expressions allow the elimination of the three integration constants ($\varphi_1$, $W_1$ and $\delta \cQ/\cQ$) and the current, $J$, to be completely expressed in terms of brane properties and $\varphi_0$.

In particular, the condition $J=0$ is satisfied when $\varphi_0 = \varphi_\star$, where
\ba \label{stability-condition}
 \left[ \left( \delta T_\ssN + \frac12 \, \delta T_\ssN'
  - \cQ \, \delta \Phi_\ssN \right)
 + \left( \delta T_\ssS + \frac12 \, \delta T_\ssS'
 - \cQ \, \delta \Phi_\ssS \right)
 \right]_{\varphi_0 = \varphi_\star} =0 \,.
\ea
This expression determines the stabilized value, $\varphi_0 = \varphi_\star$, as a function of the properties of the branes.

\subsubsection*{On-brane curvature}

Finally, the curvature in the on-brane directions, regarded as a function of $\varphi_0$, becomes
\ba \label{general-curvature}
 \left( \frac{\pi \alpha L^2 e^{-\varphi_0} }{\kappa^2} \right) \hat R (\varphi_0)
 &=& \frac12 \Bigl( \delta T_\ssN' + \delta T_\ssS' \Bigr)
 - \frac{2\pi\alpha}{\kappa^2} \left( \kappa^2 L^2 J e^{-\varphi_0} \right) \nn\\
 &=& \frac12 \left[\delta T_\ssN + \delta T_\ssS
 + \frac32 (\delta T_\ssN' + \delta T_\ssS')
 -\cQ( \delta \Phi_\ssN + \delta \Phi_\ssS) \right] \,.
\ea

Of particular interest is this result specialized to the value, $\varphi_0 = \varphi_\star$, that solves the field equations in the absence of the current $J$ (if such a value exists -- more about this below). The curvature evaluated at this value is the curvature predicted by the field equations for the brane geometry, and eq.~\pref{stability-condition} allows it to be written in two equivalent ways:
\be \label{general-curvature2}
 \left( \frac{\pi \alpha L^2 e^{-\varphi_\star} }{\kappa^2} \right) \hat R
 = \frac12 \Bigl( \delta T_\ssN' + \delta T_\ssS' \Bigr)_{\varphi_0 = \varphi_\star}
 = - \Bigl[\delta T_\ssN + \delta T_\ssS
 -\cQ( \delta \Phi_\ssN + \delta \Phi_\ssS) \Bigr]_{\varphi_0 = \varphi_\star} \,.
\ee
The second of these agrees precisely with the corresponding expression obtained in the nonsupersymmetric case studied in ref.~\cite{BvN}. However this is {\em not} also equal to the first equality of eq.~\pref{general-curvature2}, because in the nonsupersymmetric case eq.~\pref{stability-condition} no longer holds, being instead replaced by $\delta T_\ssN' + \delta T_\ssS' = 0$.

{}From the point of view of a brane observer this must agree with the (maximally symmetric) curvature that is predicted by the 4D Einstein equations given a 4D vacuum energy, $\vaceng$:
\be
 \hat R = -  4\kappa_4^2 \, \vaceng \,,
\ee
where $\kappa_4$ is the 4D gravitational coupling, given in terms of the 6D coupling, $\kappa$, by
\be
 \frac1{\kappa_4^2} = \frac{4\pi\alpha L^2 e^{-\varphi_\star}}{\kappa^2} \,.
\ee
Comparison gives
\ba
 \vaceng = - \frac{\hat R}{4\kappa_4^2} &=& - \left( \frac{\pi \alpha L^2}{
 \kappa^2} \right) \hat R \nn\\
 &=& - \frac12 \Bigl( \delta T_\ssN' + \delta T_\ssS'
 \Bigr)_{\varphi_0 = \varphi_\star} \,.
\ea
Notice that this agrees (to linear order) with the more general exact classical result obtained in eq.~(3.81) of ref.~\cite{BBvN},
\be
 \vaceng = - \sum_b \left( U_b + \frac12 \, T'_b \right) \,,
\ee
given that $U_b$ vanishes to linear order.

\subsection{The low-energy 4D effective theory}

This section constructs the effective 4D theory that reproduces the low-energy dynamics of $\varphi_0$ and the 4D metric predicted by the full 6D theory. We do so at the purely classical level, working perturbatively about a rugby ball solution, as above.

\subsubsection*{General form}

In this section the two fields of interest in the low-energy theory are the 4D metric, $\hat g_{\mu\nu}$, describing the massless KK graviton, and a 4D scalar,\footnote{We use $\varphi$ to denote the 4D field in the effective theory, to distinguish it from the (closely related) parameter $\varphi_0$ appearing in the 6D solutions.} $\varphi$, describing the low-energy would-be zero mode, $\varphi_0$, associated with the scaling symmetry of the bulk field equations. (We ignore here any other low-energy fields, such as other 4D scalars or 4D gauge fields coming from $\cA_\ssM$ or the metric.)

The most general possible local 4D effective theory describing the interactions of $\varphi$ and $\hat g_{\mu\nu}$, up to the two-derivative level, is
\be
 S_\eff = - \int \exd^4x \sqrt{-\hat g} \; \left\{
 \, \hat g^{\mu\nu} \left[ f({\varphi}) \hat R_{\mu\nu}
 + h(\varphi) \, \pd_\mu \varphi \, \pd_\nu \varphi \right]
 + V_\JF(\varphi) + j \, k(\varphi) \right\} \,,
\ee
where $f$, $h$, $V_\JF$ and $k$ are all functions to be determined, and $j$ denotes a low-energy current that is included to explore the shape of these functions (in precisely the same manner as $J$ was included in the 6D theory). Our task is to identify these functions by matching the predictions of this theory with the low-energy predictions of the full 6D system.

The functions $f$, $h$ and $k$ differ from $V_\JF$ in that they already receive their leading contributions when the two source branes are described by their background tensions, $T$; without the symmetry-breaking, $\phi$-dependent contributions $\delta T_b(\phi)$. These leading contributions can be obtained by simple dimensional reduction, which predicts
\ba
 f(\varphi) = h(\varphi) &=& \frac{4\pi \alpha L^2 e^{-\varphi}}{\kappa^2}
 = \frac{1}{2 \kappa_4^2} \, e^{-(\varphi - \varphi_\star)} \nn\\
 \hbox{and} \quad
 k(\varphi) &=& e^{-\varphi} \quad \hbox{if we define} \quad
 j \propto 4\pi\alpha L^2 J \,.
\ea
To the same approximation the (Jordan frame) potential vanishes, $V_\JF(\varphi) = 0$, since the background branes do not break the classical bulk scaling symmetry.

\subsubsection*{Low-energy matching conditions}

The goal is to determine how these quantities are perturbed by the addition of $\phi$-dependence to the brane action, $\delta T_b(\varphi)$. Our main focus is on the contribution to $V_\JF$, since (unlike for the other functions) for $V_\JF$ this is the dominant contribution. We use the prediction for the low-energy scalar curvature, $\hat R$, as a function of $\varphi$ --- {\em i.e.} eq.~\pref{general-curvature} --- as our means for doing so.

To make the comparison we compute $\hat R$ in the low-energy effective theory, assuming a maximally symmetric geometry. Defining for notational convenience $1/\hat \kappa_4^2 := e^{\varphi_\star}/\kappa_4^2 = 4\pi \alpha L^2/\kappa^2$, the metric and scalar equations of motion are
\ba \label{jRvsphi}
 e^{-\varphi} \frac{\hat R}{4 \hat\kappa_4^2}
 + je^{-\varphi} + V_\JF(\varphi) = 0\nn\\
 -e^{-\varphi} \frac{\hat R}{2 \hat\kappa_4^2} - je^{-\varphi}
 + V_\JF'(\varphi) = 0 \,.
\ea
Eliminating the current between these two equations gives the following expression for $\hat R$ as a function of $\varphi$,
\be \label{general-curvature4D}
 e^{-\varphi} \frac{\hat R}{4 \hat \kappa_4^2} = V_\JF + V_\JF'
 = e^{-\varphi} \frac{\exd}{\exd\varphi} \Bigl(e^{\varphi} \; V_\JF \Bigr) \,.
\ee

To obtain $V_\JF$ we regard eq.~\pref{general-curvature4D} as a differential equation to be integrated with respect to $\varphi$, using eq.~\pref{general-curvature} to evaluate the left-hand side as an explicit function of $\varphi$. The integral yields
\ba \label{jordan-potential-general}
 V_\JF(\varphi) &=& \frac12 \, e^{-\varphi} \int \exd\varphi_0 \;
 e^{\varphi_0} \left( \delta T_\ssN + \delta T_\ssS
 - \cQ \, \delta \Phi_\ssN - \cQ \, \delta \Phi_\ssS
 + \frac32 \, \delta T_\ssN' + \frac32 \, \delta T_\ssS' \right) \\
 &=& \frac12 \Bigl( \delta T_\ssN + \delta T_\ssS \Bigr)
 + \frac12 \, e^{-\varphi} \int \exd \varphi_0 \, e^{\varphi_0}
 \left( -\cQ\, \delta \Phi_\ssN - \cQ \, \delta \Phi_\ssS + \frac12 \, \delta T_\ssN'
 + \frac12 \, \delta T_\ssS' \right) \,. \nn
\ea
The integration constant, $C$, implicit in this integration contributes an amount $C \, e^{-\varphi}$ to $V_\JF$, with $C$ fixed by matching to the 6D theory at a specific value of $\varphi$. A convenient place for doing so is the vacuum configuration (if this exists), $\varphi = \varphi_\star$, defined by $j(\varphi_\star) = 0$, for which a prediction --- eq.~\pref{stability-condition} --- is known in the 6D theory.

Specifically, solving eqs.~\pref{jRvsphi} for $j(\varphi)$ gives
\be
 j e^{-\varphi} = -\Bigl[ V_\JF'(\varphi) + 2 V_\JF(\varphi) \Bigr] \,,
\ee
and so $\varphi_\star$ satisfies
\be
 V_\JF'(\varphi_\star) + 2 V_\JF(\varphi_\star) = 0 \,.
\ee
This has a simple interpretation in the Einstein frame, which is defined by rescaling $\hat g_{\mu\nu} = e^{(\varphi - \varphi_\star)} g_{\mu\nu}$, so that the 4D action has a canonical Einstein-Hilbert term
\be
 S_\eff = - \int \exd^4x \sqrt{-g} \; \left\{ \frac{1}{2 \kappa_4^2}
 \,  g^{\mu\nu} \left[ R_{\mu\nu}
 + 5 \, \pd_\mu \varphi \, \pd_\nu \varphi \right]
 + V_\EF(\varphi) + j_\EF \, e^{\varphi} \right\} \,,
\ee
with $j_\EF := j e^{-2\varphi_\star}$ and
\be
 V_\EF(\varphi) := e^{2( \varphi - \varphi_\star)} \, V_\JF(\varphi) \,.
\ee
Clearly $\varphi_\star$ therefore satisfies $V_\EF'(\varphi_\star) = 0$, as might have been expected. Imposing $V_\JF' + 2 V_\JF = 0$ when $\varphi = \varphi_\star$ satisfies eq.~\pref{stability-condition} then gives
\ba \label{VJFwithphistar}
 V_\JF(\varphi) &=& \frac12 \sum_b \delta T_b(\varphi) - \frac12 \, e^{-(\varphi - \varphi_\star)} \sum_b \left[ \frac12 \, \delta T_b'(\varphi_\star) +
 \cQ \, \delta \Phi_b(\varphi_\star) \right] \nn\\
 && \qquad\qquad + \frac12 \, e^{-\varphi} \int_{\varphi_\star}^\varphi
 \exd \varphi_0 \, e^{\varphi_0} \sum_b \left[ \frac12 \, \delta T_b'(\varphi_0)
 - \cQ \,\delta \Phi_b(\varphi_0) \right] \,.
\ea

Given the Einstein-frame potential, classical vacuum energy is
\ba \label{vacengexp}
 \vaceng = V_\EF(\varphi_\star) = V_\JF(\varphi_\star)
  &=& \sum_b \Bigl[ \delta T_b (\varphi_\star)
  - \cQ \, \delta \Phi_b (\varphi_\star)  \Bigr] \nn\\
  &=& - \frac12 \sum_b \delta T_b'(\varphi_\star) \,,
\ea
as found earlier (using eq.~\pref{stability-condition}). The scalar mass similarly is
\ba \label{phimassexp}
 m_\varphi^2 &=& \frac{\kappa_4^2}5 \, V_\EF''(\varphi_\star)
 = \frac{\kappa_4^2}5 \, \Bigl[ V_\JF''(\varphi_\star)
 - 4 V_\JF(\varphi_\star) \Bigr] \nn\\
 &=&\frac{\kappa_4^2}5 \sum_b \left[ \frac34 \, \delta T_b''(\varphi_\star)
 + \frac32 \, \delta T_b'(\varphi_\star)
 - \frac12 \, \cQ \,\delta \Phi_b'(\varphi_\star) \right] \,.
\ea

Similarly, chasing through the earlier expressions for the shape of the bulk geometry gives
\ba \label{geometryexp}
 \frac{\delta\alpha_b} \alpha &=& - \frac{\kappa^2}{2 \pi \alpha} \;
 \delta T_b(\varphi_\star)  \nn\\
 \frac{\delta L}L &=& - \frac{3\kappa^2}{16\pi\alpha}
 \sum_b \delta T_b'(\varphi_\star) =
 \frac{3\kappa^2  \vaceng}{8\pi\alpha}  \nn\\
 \delta W_\ssN - \delta W_\ssS &=& \frac{\kappa^2}{4\pi\alpha}
 \left[ \left( \delta T_\ssS + \frac12 \, \delta
 T_\ssS' \right) - \left( \delta T_\ssN + \frac12 \, \delta T_\ssN' \right)
 \right]_{\varphi_\star} \nn\\
 &=& \frac{\kappa^2}{8\pi\alpha} \Bigl[ \cQ \, \delta \Phi_\ssN(\varphi_\star)
 - \cQ \, \delta \Phi_\ssS(\varphi_\star) \Bigr] \,.
\ea
Notice in particular that no warping arises unless the two branes carry different amounts of localized flux. This is by contrast with the nonsupersymmetric case \cite{BvN}, for which net warping always accompanies a tension difference for the two source branes. But in the supersymmetric case the flux quantization condition does not allow such a tension difference without some of the flux being forced onto the branes.

\section{Illustrative examples}

The previous formulae with which the previous section closed represent the main results of this paper. We now explore their consequences through a number of illustrative special choices for the $\varphi$-dependence of the tensions on each brane.

\subsection{Dilaton-independent tensions and fluxes}

Consider first the simplest example: where both quantities $\delta \tau_b$ and $\delta \Phi_b$ are independent of $\varphi$. In this case the condition, $J(\varphi_\star) = 0$, defining $\varphi_\star$ degenerates to
\be
 \sum_b \Bigl(\delta T_b - \cQ \, \delta \Phi_b \Bigr) = \sum_b
 \Bigl(\delta  \tau_b - 2 \, \cQ \, \delta \Phi_b \Bigr) = 0 \,,
\ee
so two situations need to be distinguished. Either a solution to the condition $J=0$ exists --- which requires $\sum_b \delta \tau_b = 2\cQ \sum_b \delta \Phi_b$ --- or it does not. Consider each of these in turn.

\subsubsection*{When $J=0$ has solutions}

If the constant quantities $\delta \tau_b$ and $\delta \Phi_b$ satisfy the condition $\sum_b \delta \tau_b = 2\cQ \sum_b \delta \Phi_b$, then maximally symmetric solutions to the 6D field equations exist for any value of $\varphi_\star$. Because no particular value of $\varphi_0$ is selected, this shows that the flat direction that $\varphi_0$ parameterizes is not lifted. This is consistent with the observation that the brane action scales the same way as does the bulk action --- and so does not break the bulk scaling symmetry --- in the special case where $\delta \tau_b$ and $\delta \Phi_b$ are both $\varphi$-independent.

In this case formulae \pref{vacengexp}, \pref{phimassexp} and \pref{geometryexp} degenerate to $\vaceng = m_\varphi^2 = {\delta L}/L = 0$, while eqs.~\pref{geometryexp} reveal $\delta \alpha_b =\kappa^2 \delta T_b/2\pi \alpha$, as usual, and $\delta W_\ssN - \delta W_\ssS = \kappa^2 (\delta T_\ssN - \delta T_\ssS)/8\pi \alpha$. The new perturbed solution in this case is a special instance of the general solution to the full nonlinear equations \cite{GGP,GGPplus,6DdS}, all of which are known for the symmetries of interest to us. In particular, the assumption of constant brane action, $\delta T_b'=0$, is known to be sufficient to ensure $\vaceng = 0$, while $\delta T_\ssN \ne \delta T_\ssS$ induces warping. As initially argued in \cite{Towards}, it is the freedom to have nonzero on-brane flux, $\Phi_b$, that prevents the flux quantization condition from being an obstruction to reaching these solutions as perturbations to the initial rugby ball (as one might naively have thought \cite{GP}, if eq.~\pref{rugbyfluxquant} were read as forbidding the possibility of having perturbations to $\alpha_b$, and hence also to $T_b$).

\subsubsection*{When $J \ne 0$ cannot be avoided}

The perturbative solution found here also allows an exploration of what happens in the more general situation where the fluxes and tensions are {\em not} related to one another by $\sum_b \delta \tau_b = 2 \cQ \sum_b \delta \Phi_b$. In this case there is no choice for $\varphi_0 = \varphi_\star$ that can ensure $J(\varphi_\star) = 0$, implying that no solution exists at all to the linearized field equations, subject to the assumed axial symmetry and on-brane maximal symmetry. In this case studies of linearized stability \cite{stability} and exact time-dependent solutions \cite{TimeDep} suggest that the relevant solutions are necessarily time-dependent.

We now show how this expectation for time-dependence can be made more precise in the present context, since $J \ne 0$ implies the absence of a stationary point to the (Einstein-frame) scalar potential, $V_\EF(\varphi)$, for any finite value of $\varphi$. To show this we must reconsider the expression derived above for $V_\JF$, but without using the condition $V_\EF'(\varphi_\star) = 0$ to fix integration constants. For $\phi$-independent $\delta \tau_b$ and $\delta \Phi_b$ expression \pref{jordan-potential-general} for $V_\JF$ becomes
\ba \label{jordan-potential-const}
 V_\JF(\varphi) &=& \frac12 \, e^{-\varphi} \int \exd\varphi_0 \;
 e^{\varphi_0} \left( \delta T_\ssN + \delta T_\ssS - \cQ \, \delta \Phi_\ssN
 - \cQ \, \delta \Phi_\ssS + \frac32 \, \delta T_\ssN'
 + \frac32 \, \delta T_\ssS' \right) \\
 &=& \frac12 \sum_b \Bigl( \delta T_b -\cQ \, \delta \Phi_b \Bigr)
 + C \, e^{-\varphi}  \,, \nn
\ea
where $C$ is the integration constant in question.

A natural choice for $C$ is to demand that $V_\JF$ remain bounded as $e^\varphi \to 0$, since this corresponds to the weak-coupling limit for which both $\phi$ and $\cA_\ssM$ do not strongly self-interact in the bulk. More precisely, inspection of the 6D action, eq.~\pref{BulkAction}, shows that the bulk scalar potential vanishes in this limit, allowing the constant part of $\phi$ to be absorbed into the definition $\tilde\cA_\ssM := e^{-\varphi_0/2} \cA_\ssM$. This argues that $V_\JF$ should not become unbounded in this limit, leading to the requirement $C = 0$.

With this choice the Einstein-frame scalar potential becomes

\be \label{einstein-potential-const}
 V_\EF(\varphi) \propto  e^{2 \varphi} \sum_b \Bigl( \delta T_b
 - \cQ \, \delta \Phi_b \Bigr)
 =  e^{2 \varphi} \sum_b \Bigl( \delta \tau_b -2 \cQ \, \delta \Phi_b \Bigr)\,,
\ee
which describes a runaway to $\varphi \to \pm \infty$ --- whose sign depends on the sign of $\sum_b(\delta T_b - \cQ \, \delta \Phi_b)$. The absence of a solution here to $V_\EF' = 0$ for any finite value of $\varphi$ is what underlies the need for a time-dependent solution from the perspective of the low-energy 4D observer.

\subsection{Dilaton-brane couplings, vacuum energy and volume stabilization}

The next paragraphs explore some of the implications of nontrivial brane-dilaton couplings. Of particular interest is how the bulk and brane geometries depend on the choices made for these couplings. We start with the case where $\delta \tau_b$ and $\delta \Phi_b$ vary only weakly with $\varphi$, and move on to more strongly varying examples.

\subsubsection*{Linear dilaton-dependence}

Consider therefore the simple situation where both brane tensions and fluxes are linear in $\varphi$, with
\be
  \tau_b = \tau_{b0} + \tau_{b1} \, \varphi
 \quad \hbox{and} \quad
  \Phi_b = \Phi_{b0} + \Phi_{b1} \, \varphi \,,
\ee
with $\tau_{bi}$ and $\Phi_{bi}$ constant. Since many --- though not all --- physical quantities depend only on the average brane action and flux, $T_\eff := \frac12 \sum_b T_b$ and $\Phi_\eff := \frac12 \sum_b \Phi_b$, it is useful to phrase our assumptions in terms of these, which have the form
\be
 T_\eff(\varphi) = T_0 + T_1 \varphi
 \quad \hbox{and} \quad
 \Phi_\eff (\varphi) = \Phi_0 + \Phi_1 \varphi \,,
\ee
where
\be
 T_i := \frac12 \sum_{b=\ssN,\ssS} \left( \tau_{bi} - \cQ \, \Phi_{bi} \right)
 \quad \hbox{and} \quad
 \Phi_i := \frac{1}{ 2}  \sum_{b=\ssN,\ssS} \Phi_{bi}   \,.
\ee

We describe the resulting geometry as a perturbation about a rugby ball solution, characterized by a background tension, $T$, and brane flux, $\Phi(T)$, related by the background flux-quantization condition, eq.~\pref{rugbyfluxquantwbranes},
\be \label{TPhireln}
   T - \cQ \, \Phi =  \frac{2\pi}{\kappa^2 } \left[ 1 -
   \left( \frac{n g_\ssR}{g} \right) \right] \,.
\ee
With this choice, the condition $J = 0$ defining $\varphi_\star$ becomes
\ba
 0 &=& \delta T_\eff(\varphi_\star) - \cQ \, \delta \Phi_\eff (\varphi_\star)
  + \frac12 \, \delta T_\eff' (\varphi_\star) \nn\\
 &=& (T_0 - \cQ \, \Phi_0) - (T - \cQ \, \Phi) + (T_1 - \cQ \, \Phi_1) \varphi_\star
 + \frac{ T_1}{2}  \,,
\ea
whose solution,
\be \label{linearphistar}
 \varphi_\star = \frac{1}{\cQ \, \Phi_1 - T_1} \left[ (T_0 - \cQ \, \Phi_0)
 - (T - \cQ \, \Phi) + \frac{T_1}{2} \right] \,,
\ee
in this case exists so long as $T_1 \ne \cQ \, \Phi_1$.

We remark in passing that the assumed linear coupling does not preclude the existence of a vacuum configuration, $\varphi = \varphi_\star$, contrary to what happens for the nonsupersymmetric situation described in ref.~\cite{BvN}. What is different in the nonsupersymmetric case is that $\varphi_\star$ satisfies $\sum_b \delta T_b'(\varphi_\star) = 0$ --- rather than $\sum_b \left( \delta T_b + \frac12 \, \delta T_b' - \cQ \, \delta \Phi_b \right)_{\varphi = \varphi_\star} = 0$ --- which has no solutions if $\sum_b \delta T_b(\varphi)$ is a linear function of $\varphi$.

The Jordan-frame scalar potential, eq.~\pref{VJFwithphistar}, in the 4D effective theory then takes the simple form
\be
 V_\JF(\varphi) = \cQ \, \Phi_1 + (T_1 - \cQ \, \Phi_1) (\varphi - \varphi_\star)
  - (T_1 + \cQ \, \Phi_1) e^{-(\varphi-\varphi_\star)} \,,
\ee
and so the Einstein-frame potential becomes
\be
 V_\EF(\varphi) = \Bigl[ \cQ \, \Phi_1 + (T_1 - \cQ \, \Phi_1)
 (\varphi - \varphi_\star) \Bigr] e^{2(\varphi - \varphi_\star)}
  - (T_1 + \cQ \, \Phi_1) e^{(\varphi-\varphi_\star)}  \,.
\ee
Requiring the potential to be bounded from below implies $T_1 > \cQ \, \Phi_1$. Notice that at $\varphi = \varphi_\star$ this satisfies $V_\EF'(\varphi_\star) = 0$ automatically (by construction), and $V_\EF(\varphi_\star) = - T_1$ there --- which agrees with $- \frac12 \sum_b \delta T_b'(\varphi_\star) = - \delta T_\eff'(\varphi_\star)$, as it must. The physical parameters computed from $V_\EF$ using eqs.~\pref{vacengexp} and \pref{phimassexp} in this case therefore are
\be \label{linearvaceng}
 \vaceng = - T_1  \quad \hbox{and} \quad
  m_\phi^2 = \frac{\kappa_4^2}5 \, (3 T_1 - \cQ \, \Phi_1) \,,
\ee
while the extra-dimensional response of eqs.~\pref{geometryexp} becomes
\ba
 \frac{\delta\alpha_b} \alpha &=& - \left( \frac{\kappa^2}{2 \pi \alpha}
 \right) \delta T_b(\varphi_\star)
 = - \frac{\kappa^2}{2 \pi \alpha} \Bigl[ \tau_{b}(\varphi_\star)
 - \cQ \, \Phi_{b} (\varphi_\star) \Bigr] \nn\\
 \frac{\delta L} L &=& \frac{3 \kappa^2 \vaceng}{8 \pi \alpha}
 = - \frac{3 \kappa^2 T_1}{8 \pi \alpha} \,,
\ea
and
\be
  W_\ssN - W_\ssS  = \frac{\kappa^2}{8\pi\alpha} \Bigl[ \cQ \,
  \Phi_\ssN(\varphi_\star)
 - \cQ \, \Phi_\ssS(\varphi_\star) \Bigr] \,.
\ee
For potentials that are bounded from below --- {\em i.e.} those with $T_1 > \cQ \, \Phi_1$ --- the condition $T_1 > 0$ suffices to ensure $m_\varphi^2 > 0$ (and $\vaceng < 0$).

Three important properties of these expressions bear special emphasis.

First, $\vaceng$ quite generally depends on the background quantities $T$ and $\Phi$ only through the combination $T - \cQ \, \Phi$ whose value is constrained by flux quantization, eq.~\pref{TPhireln}. Consequently $\vaceng$ does not change at all as $T$ is varied, because flux quantization demands $\Phi$ must also be adjusted in a way that precisely compensates. Any value of $T$ is equally good, and what counts for physical predictions is only the extent to which the values $T_\eff(\varphi_\star) - \cQ \, \Phi_\eff(\varphi_\star)$ differ from the flux-constrained background combination, $T - \cQ \, \Phi$. This property also remains true for the more complicated examples discussed below.

Second, it is relatively easy to arrange $\varphi_\star \simeq - 50$ using only a mild hierarchy of parameters on the branes. But eq.~\pref{XDvolume} then ensures that the volume of the extra dimensions, $\cV_2 = 4\pi \alpha L^2 e^{-2\varphi_\star}$, is exponentially large compared with the intrinsic scales on the branes and in the bulk.

Third, what is most striking about this example is that the size of $\vaceng$ and $m_\varphi^2$ is completely independent of $T$, $\cQ \, \Phi$, $T_0$ and $\cQ \, \Phi_0$. In this way this example captures part of the more general magic of codimension-2 constructions; they can admit classical solutions --- like the rugby ball itself --- for which large tensions coexist with flat (or weakly curved) on-brane geometries. Why is the result independent of the $\varphi$-independent part of $T_\eff$ and $\Phi_\eff$? Quite generally, we know from eq.~\pref{vacengexp} that $\vaceng = \sum_b [\delta T_b(\varphi_\star) - \cQ \, \delta \Phi_b(\varphi_\star)] = 2 [\delta T_\eff(\varphi_\star) - \cQ \, \Phi_\eff(\varphi_\star)]$, and so (apart for the special case where $\delta T_\eff$ cancels $\cQ \Phi_\eff$) the reason $\vaceng$ can be small even when $T_0 - \cQ \, \Phi_0$ is large is because the condition $J = 0$ drives $\varphi_\star$ out to such large values that the terms $T_0 - \cQ \, \Phi_0$ and $(T_1 - \cQ \, \Phi_1) \varphi_\star$ mostly cancel in $\vaceng$.

One is drawn from this last observation to try to identify how robust this property is, both to the shape assumed for $\delta \tau_b(\varphi)$ and to the size of radiative corrections.

\subsubsection*{Power-law brane actions}

In the previous example $|\varphi_\star|$ becomes very big if $T_0$ and $\cQ\, \Phi_0$ are much larger in magnitude than are $T_1$ and $\cQ\, \Phi_1$, and so the assumption that $\delta \tau_b$ is linear in $\varphi$ typically cannot be justified simply as the first term in a Taylor expansion. It is useful therefore to examine slightly more complicated functional forms for $\delta \tau_b(\varphi)$ and $\delta \Phi_b(\varphi)$ in order to probe the robustness of the previous example.

Let us consider branes of the general form
\be
 \tau_b = \tau_{b0} + \tau_{b\eta} \, \varphi^\eta
 \quad {\rm and} \quad
 \Phi_b = \Phi_{b0} + \Phi_{b\eta} \, \varphi^\eta \, ,
\ee
again with constant $\tau_{bi}$ and $\Phi_{bi}$. The effective brane action and flux, defined as before by $T_\eff := \frac12 \sum_b T_b$ and $\Phi_b := \frac12 \sum_b \Phi_b$, then give
\be
 T_\eff(\varphi) = T_0 + T_\eta \, \varphi^\eta
 \quad \hbox{and} \quad
 \Phi_\eff (\varphi) = \Phi_0 + \Phi_\eta \, \varphi^\eta \,,
\ee
where
\be
 T_i := \frac12 \sum_{b=\ssN,\ssS} \left( \tau_{bi} - \cQ \, \Phi_{bi} \right)
 \quad \hbox{and} \quad
 \Phi_i := \frac{1}{ 2}  \sum_{b=\ssN,\ssS} \Phi_{bi}   \,.
\ee
As before we perturb about a rugby ball solution with background tension, $T$, and brane flux, $\Phi$, related by the background flux-quantization condition, eq.~\pref{TPhireln}, and so
\be
 \delta T_\eff = (T_0 - T) + T_\eta \, \varphi^\eta
 \quad \hbox{and} \quad
 \cQ \, \delta \Phi_\eff = \cQ ( \Phi_0 - \Phi) +
 \cQ \, \Phi_\eta \, \varphi^\eta\,.
\ee

We find $\varphi_\star$ by using the condition $J(\varphi_\star) = 0$, or $\delta T_\eff - \cQ \, \delta \Phi_\eff + \frac12 \delta T_\eff' = 0$, which in the present case gives
\be
 (T_0 - \cQ \, \Phi_0) - (T - \cQ\, \Phi)
 + \varphi_\star^{\eta-1} \Bigl[ ( T_\eta - \cQ\, \Phi_\eta) \, \varphi_\star
 + \frac\eta2 \, T_\eta \Bigr]  = 0 \,.
\ee
Approximate solutions are possible when $|(T_\eta - \cQ \, \Phi_\eta) \varphi_\star| \gg |\eta T_\eta /2 |$, in which case the field stabilizes approximately at
\be \label{powerphistar}
 \varphi_\star = \( \frac{D}{T_\eta - \cQ \, \Phi_\eta} \)^{1/\eta} \,,
\ee
where $D$ is defined by
\be
 D = \cQ ( \Phi_0 - \Phi) - (T_0 - T) := - (\delta T_0 - \cQ \, \delta \Phi_0) \,.
\ee
This is a real solution if the signs of $D$ and $T_\eta - \cQ \, \Phi_\eta$ are the same. For $\eta > 0$ it is also large --- and so justifies {\em a posteriori} making the large-$\varphi_\star$ approximation --- if $\abs D  \gg \abs{T_\eta - \cQ \,\Phi_\eta}$.
With this solution we find the low-energy cosmological constant is
\be \label{powervaceng}
 \vaceng = - \delta T_\eff'(\varphi_\star) =
 - \eta T_\eta \varphi_\star^{\eta - 1}
 = - \frac{\eta  T_\eta D}{T_\eta - \cQ \, \Phi_\eta}
 \( \frac{T_\eta - \cQ\, \Phi_\eta} D \)^{1/\eta} \,.
\ee
This reduces to the cases previously considered in the special cases $\eta = 0$ and $\eta = 1$. Writing $\vaceng = - (\eta D/\varphi_\star)/(1 - \cQ\, \Phi_\eta/T_\eta)$ shows this result is generically suppressed relative to $D$ within the approximations used, since these include $|\varphi_\star| \gg 1$. Because $\varphi \propto [D/(T_\eta - \cQ\, \Phi_\eta)]^{1/\eta}$, with all other things equal this suppression becomes stronger for smaller $\eta > 0$.

\subsubsection*{Exponential branes}

As our final example, consider several commonly occurring cases where the brane action depends exponentially on $\varphi$. A simple case of this type is when the entire tension and flux --- {\em i.e.} both background and perturbation --- involve a common exponential, $\tau_b (\varphi) = \tau_{b0} + \cA_b \, e^{a \varphi}$ and $\cQ \, \Phi_b (\varphi) = \cQ\, \Phi_{b0} + \cB_b \, e^{a \varphi}$.

In this case the average brane action and flux, $T_\eff := \frac12 \sum_b T_b$ and $\Phi_\eff := \frac12 \sum_b \Phi_b$, have the form
\be
 T_\eff(\varphi) = T_0 + \cA \, e^{a \varphi}
 \quad \hbox{and} \quad
 \cQ \,\Phi_\eff (\varphi) = \cQ\,\Phi_0 + \cB \, e^{a \varphi} \,,
\ee
where
\ba
 && T_0 = \frac12 \sum_b \left( \tau_{b0} - \cQ \, \Phi_{0b} \right) \,, \quad
 \Phi_0 = \frac12 \sum_b  \Phi_{0b}  \,, \nn\\
 && \cA = \frac12 \sum_b \left( \cA_{b} - \cB_{b} \right)
  \quad \hbox{and} \quad
 \cB  = \frac12  \sum_b \cB_{b}  \,.
\ea
For instance, given the explicit factor of $e^{-\phi}$ in the definition of the brane-flux coupling, eq.~\pref{BraneFluxCoupling}, the special case $a = 1$ and $\cA_b = \Phi_{b0} = 0$ (and so $\Phi_0 = \cA + \cB = 0$) corresponds to having brane actions that do not directly couple to the bulk scalar $\phi$.

As before we perturb about a rugby ball solution with background tension, $T$, and brane flux, $\Phi$, related by the background flux-quantization condition, eq.~\pref{rugbyfluxquantwbranes},
\be \label{TPhirelnagain}
   T - \cQ \, \Phi = \frac{2\pi}{\kappa^2} \left(
    1 - \frac{n g_\ssR}{g} \right) \,.
\ee
The perturbations about this background become
\be
 \delta T_\eff = (T_0 - T) + \cA \, e^{a \varphi}
 \quad \hbox{and} \quad
 \cQ \, \delta \Phi_\eff = \cQ ( \Phi_0 - \Phi) + \cB \, e^{a \varphi}\,.
\ee
The condition $J(\varphi_\star) = 0$ defining $\varphi_\star$ as usual is $\delta T_\eff (\varphi_\star) + \frac12 \, \delta T_\eff'(\varphi_\star) - \cQ \, \delta \Phi_\eff(\varphi_\star) = 0$, which in this case becomes
\be
 \left[ \cA \left( 1+\frac a2 \right) - \cB \right]
 e^{a \varphi_\star}  = D \,,
\ee
where $D := \cQ ( \Phi_0 - \Phi) - (T_0 - T) $. This has solutions if the sign of both sides is the same.

The low-energy Jordan-frame potential, eq.~\pref{VJFwithphistar}, then is
\ba
 V_\JF(\varphi) &=& C_1 e^{a \varphi} + C_2 e^{-\varphi} + C_3 \,, \\
 \hbox{with} \quad C_1 &=& \frac{1}{a+1} \left[
 \left(1 +\frac{3a}{2} \right) \cA - \cB \right]\nn\\
  C_2 &=& \left[\frac 2{\(\cA+\frac12 a \cA-\cB\)^{1/a}}
  -\frac{(2+a)C_1}{\(\cA+\frac12 a\cA-\cB\)^{1+1/a}}  \right]
  D^{1+1/a} \nn\\
  \hbox{and} \quad C_3 &=& -D \,, \nn
\ea
leading to a similar expression for the Einstein-frame potential, $V_\EF = V_\JF \, e^{2(\varphi - \varphi_\star)}$. At $\varphi = \varphi_\star$ the cosmological constant becomes
\ba \label{vacengexponential}
 \vaceng = -  \delta T_\eff'(\varphi_\star)
 = - {a \cA} \, e^{a \varphi_\star}
 &=& a  \left[
 \frac{(T_0 - T) - \cQ (\Phi_0 - \Phi) }{ 1 + (a/2) - \cB / \cA } \right] \\
 &=& a  \left[ \frac{(T_0 - \cQ\, \Phi_0) - (2\pi/\kappa^2)
 ( 1 - n g_\ssR/g) }{ 1 + (a/2) - \cB / \cA } \right] \,, \nn
\ea
where the last equality uses the value of $T - \cQ \, \Phi$ dictated by flux quantization. The scalar mass at the extremum is similarly
\ba
 m_\varphi^2 &=& \frac{a}{10} \left(
 \frac{6 + 3a - 2\cB/\cA}{2 +a -2\cB/\cA} \right) \kappa_4^2
 \Bigl[ (T - T_0) - \cQ ( \Phi - \Phi_0) \Bigr]\nn\\
 &=& - \frac{a}{10} \left(
 \frac{6 + 3a - 2\cB/\cA}{2 +a -2\cB/\cA} \right) \kappa_4^2
 \left[  (T_0 - \cQ\, \Phi_0) - \frac{2\pi}{\kappa^2}
 \left( 1 - \frac{n g_\ssR}{g} \right) \right] \,.
\ea

Eq.~\pref{vacengexponential} identifies three potential mechanisms for suppressing $\vaceng$.
\begin{enumerate}
\item The first is if $a \to 0$, in which case the brane actions become $\varphi$-independent and $\varphi_\star$ recedes to infinity. This is the suppression already encountered in the examples presented above.
\item The second is if the $\varphi$-independent parts, $T_0$ and $\Phi_0$, are related to one another in the same way as flux quantization imposes on the background values, $T$ and $\Phi$. (For the special case where the bulk flux is chosen to lie in $U_\ssR(1)$ direction (so $g = g_\ssR$) and $n = 1$, the background lies in the same flux category as does the supersymmetric Salam-Sezgin solution \cite{SSs}, for which $T - \cQ \, \Phi = 0$.)
\item Finally, the third potential suppression occurs even if $T - \cQ \, \Phi \ne 0$, provided $|\cB / \cA| \gg 1$. That is, if $\delta \Phi_\eff$ dominates $\delta T_\eff$ then this only affects the value of $\varphi_\star$, leaving $\vaceng$, as always, of order $\delta T_\eff'(\varphi_\star)$.
\end{enumerate}

\subsection{Quantum corrections and technical naturalness}
\label{subsec:quantum}

All of the calculations of brane-bulk interactions provided in previous sections are performed purely within the classical approximation. As such they leave open the question of how robust their conclusions are to modification by quantum corrections. And since the streets are littered with classical examples having small vacuum energies, a proper treatment of quantum corrections is the crucial to any credible mechanism for understanding the small size of the observed vacuum energy.

In this section, we take a small step towards filling in this missing step, more in the spirit of indicating a promising line of inquiry than in providing a polished example. Our interest is in quantifying the stability of both the size of the low-energy cosmological constant, $\vaceng$, and the size of the bulk volume, $\cV_2$, (when this is large compared with more microscopic scales).

The starting point is an enunciation of the essence of the problem: once parameters are chosen to ensure a large value for $\vaceng$ and/or $\cV_2$, are these choices stable against the renormalization that results when heavy fields are integrated out? In extra-dimensional brane models this question necessarily has two parts, to do with integrating out heavy field on the brane and in the bulk.

We here use one of the previously discussed examples as a toy model for estimating the size of quantum corrections. We choose a model that has both has an exponentially large volume and a small 4D on-brane curvature --- {\em i.e.} vacuum energy\footnote{In the model $\vaceng$ is small inasmuch as it is parametrically suppressed relative to other scales, though not small enough numerically to describe the observed Dark Energy.} --- and estimate the size of quantum corrections. The main idea behind this model is that it is the bulk field $\phi$ itself that counts both bulk and brane loops, with weak coupling corresponding to $\phi$ being large and negative. The influence of loop effects is then simply incorporated by tracking the $\phi$-dependence of the quantum-corrected (1PI) action, for which the above arguments about brane-bulk back-reaction can be applied.

\subsection*{A toy model}

The theory of interest is one of the `power-law' models described earlier. For our starting point we take a background rugby-ball geometry whose background tension and flux satisfy the flux-quantization condition, eq.~\pref{TPhireln},
\be
 T - \cQ \, \Phi = \frac{2\pi}{\kappa^2} \left[ 1 - \left( \frac{n g_\ssR}{g}
 \right) \right] \,,
\ee
with the right-hand-side being small enough to allow semiclassical reasoning, but not tuned to be inordinately small. Such geometries have flat on-brane directions, and as above we seek to see how brane-bulk interactions modify this, including loops.

For the perturbations to this geometry we choose the classical brane-bulk Lagrangian to have the power-law form,
\be \label{toybrane}
 \delta T_\eff = T_\star \, (-\phi)^\eta \,,
\ee
with $0 < \eta < 1$ (and the smaller $\eta$ is, the larger the suppression in $\vaceng$). Here $T_\star$ is a function of all of the on-brane degrees of freedom, $\psi$, such as
\be \label{toybrane2}
 T_\star = \mu^4  +
 \hat g^{\mu\nu} \, \partial_\mu \psi \, \partial_\nu \psi
 + M^2 \psi^2 + \lambda \, \psi^4 + \cdots \,,
\ee
which defines the scale $\mu$. In the vacuum $\psi = 0$ and so $T_\star = \mu^4$.

A hierarchy is dialled in by choosing the scale $\mu$ in $T_\star$ to be small compared with the typical brane scale, $M$: {\em i.e.} $\mu^2 \ll M^2 \ll 1/\kappa$. Notice that taking $-\phi \gg 1$ does not affect the mass of the $\psi$ particle (or other brane particles in general) at the classical level, because $\phi$ appears only as an overall factor in the brane action. The goal is to show that the energy scales set by $\vaceng^{1/4}$ and $\cV_2^{-1/2}$ can be hierarchically different from $M$, and that this can be protected from quantum effects. Since $\cV_2$ turns out to depend exponentially on $T_\star$, a relatively small hierarchy between $T_\star$ and $M^4$ suffices to generate very large volumes.

The classical part of the story is worked out above, with (choosing $\Phi_\eff = 0$) eq.~\pref{powerphistar} implying
\be
 -\varphi_\star \simeq \left( \frac{T - \cQ \, \Phi}{T_\star} \right)^{1/\eta}
 \simeq \left[ \frac{2\pi}{\kappa^2 \mu^4} \left( 1 - \frac{n g_\ssR}{g} \right) \right]^{1/\eta} \,,
\ee
from which eq.~\pref{XDvolume} gives the bulk volume,
\be
 \cV_2 = 4 \pi \alpha L^2 e^{-\varphi_\star}
 \simeq 4 \pi \alpha L^2 \, \exp \left\{ \left[ \frac{2\pi}{\kappa^2 \mu^4} \left( 1 - \frac{n g_\ssR}{g} \right) \right]^{1/\eta} \right\} \,.
\ee
Eq.~\pref{powervaceng} similarly gives the on-brane vacuum energy as
\be
 \vaceng \simeq \eta T_\star (- \varphi_\star)^{\eta - 1}
 \simeq \frac{\eta (T - \cQ \, \Phi) }{(-\varphi_\star)}
 \simeq \frac{2\pi \eta}{\kappa^2} \left( 1 - \frac{n g_\ssR}{g} \right)
 \left[ \frac{2\pi}{\kappa^2 \mu^4} \left( 1 - \frac{n g_\ssR}{g} \right) \right]^{-1/\eta}\,,
\ee
revealing a power-law suppression of $\vaceng$ relative to $1/\kappa^2$, whose strength improves the smaller $\eta$ gets. ({\em e.g.} for $\eta = 1$ this gives $\vaceng \sim \mu^4$ while $\eta = \frac12$ implies $\vaceng \propto \kappa^2 \mu^8$, and so on.)

\subsubsection*{Loop corrections}

We now argue that the choice $\mu \ll M$ underlying the classical hierarchy is technically natural. We do so using the observation that it is the expectation of the bulk zero-mode, $\varphi$, itself that controls the size of these loops, so loop corrections can be incorporated into the above argument by making a modified choice for the $\phi$-dependence of $T_\eff$.

\medskip\noindent{\em Brane loops:}

\medskip\noindent
To see why this is so, imagine first computing quantum corrections involving loops of the on-brane field, $\psi$. When computing these loops it is useful first to adopt a canonical normalization, $\psi \to \psi_c := (- \varphi)^{\eta/2} \,\psi$, after which the strength of the self-coupling becomes revealed to be $\lambda_c \psi_c^4$ with
\be
 \lambda_c = \frac{\lambda }{(- \varphi)^{\eta} }  \,.
\ee

More generally, because $(- \varphi)^\eta$ pre-multiplies the entire brane action, for the purposes of power-counting brane perturbation theory it plays the role of $1/\hbar$. This ensures that each additional loop is parametrically suppressed by an additional factor of $(- \varphi)^{-\eta}$, with dimensions made up using the typical brane scale, $M$. In particular, integrating out a heavy field of mass $M$ should give a Wilson action (or, alternatively a calculation of the `quantum' 1PI brane action) of the form
\be \label{loopbrane}
 \Gamma_\eff = T_\star (-\phi)^\eta +  T_1 + \frac{T_2}{(- \phi)^{\eta}}
 + \cdots \,,
\ee
and so on.

Here the $T_n$ generically depend on the brane fields,\footnote{Although these would be local for the Wilson action, they need not be for the 1PI action \cite{EFTrev}.} much as did $T_\star$. The point of quantum hierarchy problems is that --- on dimensional\footnote{These are often stated to be of order the `cutoff' scale, but we make the more conservative statement that they scale with the physical mass $M$ because cutoffs generically cancel in all physical quantities \cite{UsesAbuses}.} grounds --- each of the $T_n$ is generically of order $M^4$, rather than the smaller $\mu^4$. The question is whether this ruins the above conclusions about the size of $\varphi_\star$, and so also of $\cV_2$ and $\vaceng$.

Now comes the main point. Because each loop correction is suppressed by an additional factor of $(-\phi)^{-\eta}$, none of them has the same $\phi$-dependence as does $T_\star$. In particular, none of them require the vacuum value of $T_\star$ also to be of order $M^4$ instead of $\mu^4$. Better yet, having $T_1 \simeq M^4 \gg T_\star \simeq \mu^4$ keeps $\varphi_\star$ stabilized at large negative values, enforcing the dominance of the leading, classical, approximation.

To see this in detail we repeat the above classical calculation of the potential for the bulk modulus $\varphi$ using the loop corrected action, eq.~\pref{loopbrane}, rather than the classical expression, eqs.~\pref{toybrane} and \pref{toybrane2}. For simplicity we take $\Phi_\eff = 0$ also at the loop level, though none of our conclusions would change if we were to assume a similar loop expansion for the brane flux,
\be
 \Phi_\eff \simeq \Phi_1 + \frac{\Phi_2}{(- \phi)^\eta} + \cdots \,,
\ee
with the dimensions of the $\Phi_n$ again set by the large mass, $M$, circulating in the loops.

To one-loop order, the new terms in $\Gamma_\eff$ are independent of $\phi$, and so their modifications to the brane-bulk back-reaction are encompassed by the analysis given in the previous section. Assuming $- \varphi_\star \gg 1$ this gives
\be
 -\varphi_\star \simeq \(\frac{D}{T_\star}\)^{1/\eta}
 \quad \hbox{and} \quad
 \vaceng \simeq \frac{\eta D}{(-\varphi_\star)} \,,
\ee
where $D = (T - T_1) - \cQ(\Phi - \Phi_1) \simeq (T - \cQ \, \Phi) - T_1$ is of order the larger of $M^4$ or $T - \cQ\, \Phi \propto 2 \pi/\kappa^2$. Since both of these scales are much larger than $T_\star \propto \mu^4$, the classical assumption that $-\varphi_\star$ is large (and all that comes with it) is not undermined by one-loop corrections. We assume here that $D$ and $T_\star$ share the same sign.

The size of the two-loop correction can be similarly estimated. The equation that determines $\varphi_\star$ is, at two-loop order
\be
 -D + T_\star (-\varphi_\star)^\eta
 \left( 1 + \frac{\eta}{2 \varphi_\star} \,
 \right) + \frac{T_{2}}{ (-\varphi_\star)^{\eta}}
 \left( 1 - \frac{\eta}{2 \varphi_\star} \,
 \right) \simeq 0 \,.
\ee
Using $|D| \sim |T_2| \sim \cO(M^4)$ and $(-\varphi_\star)^\eta \sim M^4/\mu^4 \gg 1$ shows the 2-loop term to represent a small correction to the value predicted for $(-\varphi_\star)$.

The two-loop contribution to $\vaceng$ comes in two parts. The first of these is through the change in $\varphi_\star$, though because $|\delta \varphi_\star / \varphi_\star| \ll 1$ this contribution is subdominant to the value for $\vaceng$ already computed at one loop. In particular, it doesn't ruin the suppression of $\vaceng$ by the factor $1/\varphi_\star$.

The second type of 2-loop contribution to $\vaceng$ comes from the fact that $T_\eff'$ now includes a new term of the form
\be
 -\delta T_\eff' \simeq \eta \, T_{2} (-\varphi_\star)^{-\eta-1}
 = \left( \frac{\eta T_{2}}{-\varphi_\star} \right)
 \frac{1}{(-\varphi_\star)^\eta} \,,
\ee
which should be compared to the original contribution, $\sim \eta D/ \varphi_\star$. Clearly the new term is subdominant if $|D| \sim \abs{T_{2}} \gg T_\star$. And so it goes for higher loops, each of which is suppressed by an additional factor of $(-\varphi_\star)^{-\eta} \simeq \mu^4/M^4$, by virtue of the stability of the initial stabilization at large values of $-\varphi_\star$.

Notice that the brane loops also don't have a large relative impact on the light scalar mass, which eq.~\pref{phimassexp} gives to be
\ba
 m_\varphi^2 &\simeq& \frac{\kappa_4^2}5 \left[ \frac32 \, \delta T_\eff''(\varphi_\star) + 3 \, \delta T_\eff'(\varphi_\star)
 - \cQ \,\delta \Phi_\eff'(\varphi_\star) \right] \\
 &\simeq& \left( \frac{3 \eta T_\star}{5 M_p^2} \right)
 (-\varphi_\star)^{\eta - 1} \simeq \frac{ 3 \eta D}{5 M_p^2 \varphi_\star}
 \sim \frac{\eta M^4}{M_p^2} \left( \frac{\mu}{M} \right)^{4/\eta}
 \qquad\;\, \hbox{(up to one loop)} \nn\\
 &\simeq& \left( \frac{3 \eta T_2}{5 M_p^2} \right) (-\varphi_\star)^{-\eta-1}
 \simeq \frac{3 \eta T_2 T_\star}{5 M_p^2 D \varphi_\star}
 \sim \frac{\eta \mu^4}{M_p^2} \left( \frac{\mu}{M} \right)^{4/\eta}
 \quad\;\, \hbox{(two-loop term)} \,, \nn
\ea
with the final estimates using $T_2 \sim D \sim M^4$ and $T_\star \sim \mu^4$.

\medskip\noindent{\em Bulk loops:}

\medskip\noindent
The previous loop estimates are restricted purely to brane loops because they rely on the assumed form of the brane-bulk coupling. But the back-reaction of the brane loops onto the bulk is computed classically (for the bulk theory) just as before. How big might be quantum corrections in the bulk sector?

An estimate for the size of bulk loops can be made in a manner very similar to the one just used for brane loops, because $e^{2\phi}$ is the loop-counting parameter for the bulk 6D supergravity. The simplest way to see this is to re-scale the 6D metric according to $g_{\ssM \ssN} \to \check g_{\ssM \ssN} := e^{-\phi} g_{\ssM \ssN}$, in terms of which the action of eq.~\pref{BulkAction} becomes
\be \label{BulkActionSF}
 S_\mathrm{bulk} = - \int \exd^6 x \sqrt{- \check g} \;
  e^{-2\phi} \left\{ \frac1{2\kappa^2} \, \check g^{\ssM\ssN}
 \Bigl( \check \cR_{\ssM \ssN} + \zeta \pd_\ssM \phi \, \pd_\ssN \phi \Bigr)
 + \frac14 \, \check g^{\ssM \ssP} \check g^{\ssN\ssQ}
 \cF_{\ssM\ssN} \cF_{\ssP\ssQ}
 + \frac{2 \, g_\ssR^2}{\kappa^4} \right\} \,,
\ee
where $\zeta$ is a constant. This shows that for bulk perturbation theory it is the constant value of $e^{2\phi}$ that plays the role of of $\hbar$. (The same also remains true once the action's fermion terms are included \cite{NS}.) Each loop involving bulk fields therefore contributes an amount proportional to an additional power of $e^{2\phi}$, which is small when $\phi$ is large and negative (also the regime of weak brane coupling).

Now imagine integrating out fields in the bulk that are heavy relative to the KK scale. Loops of these fields potentially modify both the brane and bulk actions by new local interactions \cite{6Dlocal,GilkeyReviews}. The loop-generated couplings arising in this way cannot depend on $\varphi$ in the same way as does the classical action, \pref{BulkAction}, again indicating that these classical terms are not themselves renormalized. Loop-generated terms necessarily involve new interactions whose $\varphi$-dependence can be organized into a series in powers of $e^{2\varphi}$ \cite{6Dbulkloop}.

This leads one to expect that each bulk loop is exponentially suppressed, by powers of $e^{2\varphi_\star} \propto 1/\cV_2^2$ when $- \varphi_\star \gg 1$. In particular, these corrections to physical properties would therefore be expected to be sub-dominant to the brane loops considered above.

\section{Conclusions}

This paper computes the back-reaction of a pair of 4D codimension-two branes onto the 6D geometry that they source, within a framework of flux compactification that allows a complete calculation of modulus stabilization. Although performed with a particular (gauged, chiral \cite{NS}) 6D supergravity, the mechanisms exposed by our calculations rely only on broad features (like the presence of a dilaton and scale invariance of the classical equations) shared by a wide variety of higher-dimensional supergravities. This leads us to expect them to have a wider domain of validity than the particular 6D system studied here.

The main calculational assumptions are these: ($i$) we assume all energy densities and curvatures to be small enough to justify working within a semi-classical analysis; ($ii$) we compute brane-bulk couplings to leading order in a derivative expansion (making the dominant players the brane tensions and brane-localized fluxes); ($iii$) we seek solutions that are axially symmetric in the two dimensions transverse to the branes and whose on-brane geometries are maximally symmetric; and ($iv$) we linearize the brane properties about the choices that source simple rugby-ball geometries (which have flat on-brane geometries despite having nonzero tensions).

The last two of these assumptions deserve some motivation. The linearization about rugby ball geometries \cite{Towards,SSs,GGP,GGPplus} is made in order to allow the search of their immediate neighborhoods in field space to be systematic; the linearity of the equations allows the construction of their most general solutions. We do not believe that the qualitative features of our results (like the existence of very large volume solutions, and the suppression of on-brane curvatures) depend strongly on this assumption.

By contrast, at first blush the assumption of maximal symmetry might seem more restrictive, since maximal symmetry is known {\em not} to be possible for a majority of brane configurations and the general situation is expected to be time-dependent \cite{TimeDep}. We employ a trick to explore such configurations: we stabilize the time-dependent runaway by turning on an external current that couples to the system's low-energy moduli. In this way we can explore the potential energy cost that drives these runaway solutions, at least at the low energies of main interest.

The supergravity of interest has a one-parameter flat direction, labeled by a particular combination of the 6D dilaton and the breathing mode of the extra-dimensional metric. Our main calculation interest for this theory is in the potential energy generated for this flat direction by the back-reaction of the bulk-brane couplings; and in the related change of shape of the extra-dimensional geometry. We find that these display the following noteworthy features:
\begin{itemize}
\item {\em Volume stabilization:} Any non-derivative coupling of the branes to the bulk dilaton, $\phi$, breaks the classical scaling symmetry of the bulk field equations, and so lifts the degeneracy of the classical zero mode. Because the bulk volume depends exponentially on the canonically normalized dilaton, $\phi$, we find that a mild hierarchy in the brane-bulk coupling parameters can easily generate an exponentially large extra-dimensional volume. The exception to this is if the branes also couple only to exponentials of $\phi$, as is in particular often true for $D$-branes.
\item {\em Suppressed on-brane curvature:} A remarkable feature of rugby ball geometries is that their on-brane directions are flat despite the presence of large brane tensions. We find that perturbations about these geometries can -- but need not -- share this feature, having on-brane curvatures that are parametrically small compared with the generic size of the on-brane tensions. In particular, a mechanism for achieving such solutions arises for some types of dilaton-brane couplings since the dilaton can be driven to roll out to large fields along the flat direction to find places where the on-brane tension and curvature are the smallest.
\item {\em Relevance of on-brane fluxes:} Flux quantization within the bulk provides a strong constraint on flux-stabilized rugby-ball geometries, and in the simplest examples gives rugby-ball perturbations whose on-brane curvatures are {\em not} suppressed. An important part of our ability to find other solutions with lower curvature is our inclusion of brane-localized flux, corresponding to a magnetic coupling of the branes to the geometry-stabilizing fluxes. In this way our calculations bear out the earlier expectations of ref.~\cite{Towards}.
\item {\em Quantum corrections:} In section \ref{subsec:quantum} we provide a preliminary estimate of the size of quantum corrections for a particularly promising toy model, with both exponentially large volumes and a suppressed on-brane curvature. Our estimates indicate these properties need not be destabilized by quantum effects on the brane or in the bulk. They do not do so because it is the value taken by the dilaton along the classical flat direction itself that plays the role of the loop-counting parameter (similar to what happens for string vacua), and this constrains how quantum effects can alter the dynamics that determines what this value is. In particular, we find that provided the brane couplings are arranged to lie within the regime of weak coupling, the conclusions of the classical analysis are protected from loop effects.
\item {\em Bulk distortion:} Although the rugby ball solutions themselves are simple we find that the nearby geometries are more generic, including warping and nontrivial dilaton profiles across the extra dimensions.
\item {\em Modulus-matter couplings:} Having the dilaton couple to branes only through an overall prefactor implies a universal coupling to ordinary matter, if this resides on a brane. This is likely to have interesting phenomenological implications if the moduli can be arranged to be light enough to mediate macroscopic forces. It is noteworthy that these couplings can have the form required to profit from a `chameleon' mechanism \cite{chameleon}.
\end{itemize}

We regard these properties to be new examples of how low-energy brane dynamics can change how one thinks about technical naturalness and hierarchies of scale. In particular, the natural generation of exponentially large volumes in the 6D model explored here fills in a key missing step in efforts to use large volumes to solve the gauge hierarchy problem.

Although none of the solutions explored here have on-brane curvatures that are low enough to describe the Dark Energy density, the existence in some models of a mechanism for robust parametric suppression of the on-brane curvature is very suggestive. We regard this as encouragement to continue to explore this direction for new approaches to the cosmological constant problem.

\section*{Acknowledgments}

We wish to thank Markus Luty, Fernando Quevedo and Raman Sundrum for useful discussions about backreaction in codimension-two models. CB acknowledges the Kavli Institute for Theoretical Physics in Santa Barbara and the Abdus Salam International Center for Theoretical Physics for providing the very pleasant environs in which some of this work was performed, as well as Eyjafjallajokull for helping to provide some unexpected but undivided research time. LvN thanks the Instituut-Lorentz for Theoretical Physics at Leiden University for their hospitality. Our research is supported in part by funds from the Natural Sciences and Engineering Research Council (NSERC) of Canada. Research at the Perimeter Institute is supported in part by the Government of Canada through Industry Canada, and by the Province of Ontario through the Ministry of Research and Information (MRI).

\appendix

\section{Flux quantization with brane fluxes}
\label{App:fluxconditionws}

To see how to interpret the parameter $\Phi_b$, rewrite the brane flux term as a regularized 6D integral weighted by a scalar function $s(\rho)$ whose support is nonzero only in a short interval $|\rho - \rho_b| < \varepsilon$ away from the brane, and is normalized so that $\int \exd^2x \, \sqrt{g_2} \; s = 1$. That is,
\ba \label{app:6Dfluxform}
 S_\mathrm{flux} &=& \frac{\Phi_b}{2} \, \int \exd^4x \,\sqrt{- g_6}e^{-\phi} \; s \,
 \epsilon^{mn} \cF_{mn}  =   \Phi_b \, \int \exd^6x \, \sqrt{-g_4}e^{-\phi}
 \; s \, \cF_{\rho\theta} \,.
\ea
Then the $\delta \cA_\theta$ Maxwell equation becomes
\be
 \partial_\rho \Bigl(e^{-\phi} \sqrt{-g_6} \; \cF^{\rho\theta} - e^{-\phi}\Phi_b \sqrt{-g_4} \; s
 \Bigr) = 0 \,,
\ee
which integrates to give
\be \label{app:maxsolnws}
 \Bigl( e^{-B} \cA_\theta' - \Phi_b s \Bigr) = \cQ e^\phi \,.
\ee
This is the bulk solution found in the text away from the brane, where $s = 0$.

Imagine now integrating this to obtain $\cA_\theta(\rho)$ in the vicinity of the brane at $\rho_b = 0$, using for $s$ a simple step function: $s = 1/(\pi \varepsilon^2)$ for $\rho < \varepsilon$ and $s = 0$ for $\rho > \varepsilon$. Assuming $W \simeq W_b$ is approximately constant and $e^B \simeq \rho$ for $\rho < \varepsilon$, we set $\cA_\theta(0)=0$ and integrate tof find $\cA_\theta(\varepsilon)$,
\be
 \cA_\theta(\varepsilon) = \frac{\Phi_b}{\pi \varepsilon^2}\left[\frac12\rho^2\right]^\varepsilon_0+\cQ\int_0^\varepsilon\exd\rho\rho e^\phi=\frac{\Phi_b}{2\pi}   +\cQ\int_0^\varepsilon\exd\rho\rho e^\phi \,,
\ee
and so as long as $e^\phi$ diverges less fast than $\rho^{-2}$ at the brane we have
\be
 \lim_{\varepsilon\rightarrow0}\cA_\theta(\varepsilon) = \frac{\Phi_b}{2 \pi} \,.
\ee

The junction condition for $\cA_\theta'$ at $\rho = \varepsilon$ can also be seen by subtracting the solution, eq.~\pref{app:maxsolnws} evaluated at $\rho < \varepsilon$ --- where $s = 1/(\pi \varepsilon^2)$ --- from the same solution evaluated at $\rho > \varepsilon$ --- where $s = 0$. Since the RHS is the same in both cases we get the following jump discontinuity across $\rho = \varepsilon$:
\be
 \Bigl[ e^{-B} \cA_\theta' \Bigr]^{\rho = \varepsilon+}_{\rho = \varepsilon-}
 = - \frac{\Phi_b}{ \pi \varepsilon^2} \,.
\ee
This can be related to the derivative of the brane action with respect to $\cA_\theta$ by rewriting eq.~\pref{app:6Dfluxform} as
\be
 S_\mathrm{flux} = \Phi_b \, \int \exd^6x \, \sqrt{-g_4}
 \; s \, \cF_{\rho\theta}  = \frac{2\pi \Phi_b}{\pi \varepsilon^2} \,
 \int \exd^4x \, \sqrt{- g_4} \; \cA_\theta(\varepsilon)
 \,,
\ee
and so (keeping in mind the relative sign between the tension and flux terms)
\be
 \Bigl[ e^{-B} \cA_\theta' \Bigr]^{\rho = \varepsilon+}_{\rho = \varepsilon-}
 = + \frac{1}{2 \pi} \; \left( \frac{\partial T_b}{\partial \cA_\theta}
 \right) \,,
\ee
as stated in ref.~\cite{BBvN}.

\section{Alternative currents}
\label{App:alternative current}

In this section, we check that the details of the current are not important for the stabilization of $\varphi$. Define for comparison purposes the current
\be
 S_\ssJ = -\int \exd^6 x \, \sqrt{-g} \; J \, e^\phi \,.
\ee
This choice keeps the scale invariance of the bulk action intact. Compared to the current used in the main body, the changes to the linearized equations of motion arise only in the $\phi$ and $\delta B$ equations. Here we write only the contributions to these equations due to the current:
\ba
 \( \sin x \, \delta \phi' \)' &=& \cdots  - \kappa^2 J L^2 e^{\varphi_0} \sin x \nn\\
 \frac{\( \sin^2 x \, \delta B' \)'}{\sin^2 x} &=& \cdots -\kappa^2 J e^{\varphi_0} \,,
\ea
where the $\delta B$ term was present before but lacked the factor $e^{\varphi_0}$. By contrast the $\delta\phi$ contribution given above didn't exist for the current used in the main text.

The resulting change to the perturbations that is a consequence of the current only is
\ba
 (\delta \phi)_\ssJ &=& \kappa^2 J L^2 e^{\varphi_0} \ln\abs{\sin x} \nn\\
 (\delta B)_\ssJ &=& \kappa^2 J L^2 e^{\varphi_0} \left[ -\cH_2(x)
 + \frac12 \( 1+ x \cot x \) \right] \,.
\ea
The resulting changes in the matching conditions to the brane are as follows:
\ba
 \frac{\delta\cQ}{\cQ} + \kappa^2 J L^2 e^{\varphi_0} &=&
 \frac{\kappa^2}{4\pi\alpha} \Bigl( T_\ssN' + T_\ssS' \Bigr) \nn\\
 \frac12 \, \kappa^2 J L^2 e^{\varphi_0} &=& -\frac{\kappa^2}{4\pi\alpha}
 \left[ \delta T_\ssN + \delta T_\ssS - \cQ\Phi_\ssN - \cQ\Phi_\ssS
 + \frac12\, ( T_\ssN' + T_\ssS' ) \right] \,.\nn\\
\ea
Relative to the main text, the current appearing in the first of these equations is new, and in the second equation it is half the size as found in the main text (apart from the trivial scaling by $e^{\varphi_0}$ throughout). The resulting 4D curvature is unchanged because it is the combination
\be
 L^2 \hat R = 2 \left( \frac{\delta\cQ}\cQ \right)
 - 2 \kappa^2 J L^2 e^{\varphi_0} = \frac{\kappa^2}{2\pi\alpha} \Bigl(
 T_\ssN' + T_\ssS' \Bigr) - 4 \kappa^2 J L^2 e^{\varphi_0} \,.
\ee
Here the current contribution is twice the result of the main text, so with the current being only half as large for a given $\varphi_0$, the curvature remains unchanged.

Using the corresponding current in the 4 dimensional theory --- {\em i.e.} using $\sqrt{-\hat g} \; j$ --- yields in general a different Einstein-frame effective potential. However, $V_\EF(\varphi_\star)$ and $V_\EF''(\varphi_\star)$ agree. Since the coefficient of the kinetic term is unchanged, neither is the mass of the dilaton and the cosmological constant. This shows that we can extract the properties at the stationary point reliably, even though the shape of the potential away from this point can depend on the detailed definition of the current that is used. This reflects a general property: the detailed form of a scalar potential can be varied (as always) by performing a field redefinition, though any dependence on the field variables used ultimately drops from any physical prediction.

\section{Linearization around the rugby ball}
\label{app:linearization}

This appendix computes the linearization of the field equations about the rugby ball solutions, with
\ba \label{app:linearizationdefs}
 e^B &=& e^{B_0}(1 + \delta B) = e^{-\varphi_0/2}
 \alpha L \sin \left( \frac {\hat\rho} L \right) (1+\delta B)\nn\\
 W &=& \delta W \qquad \hbox{and} \qquad
 \phi = \varphi_0 + \delta\phi \\
 \cF_{\rho\theta} &=& ( \cQ + \delta\cQ ) e^{\phi + B - 4W} =
 \cQ e^{\varphi_0 + B_0} \left( 1 + \frac{\delta\cQ}{\cQ}
 + \delta B + \delta\phi - 4 \delta W \right) \,. \nn
\ea

Using these in the d'Alembertian for the dilaton gives
\ba
 \Box \phi &=& \frac1{\sqrt{-g}} \, \pd_\ssM \left( \sqrt{-g} \, g^{\ssM \ssN}
 \pd_\ssN \phi \right)
 = \frac{e^{\varphi_0}}{\sqrt{-g}} \, \partial_{\hat \rho} \left( \sqrt{-g} \,
 \partial_{\hat \rho} \phi \right) \nn\\
 &=& \frac{e^{\varphi_0}}{\sin\left(\hat\rho/ L\right)} \,
 \partial_{\hat \rho} \left[ \sin \left( \frac {\hat\rho} L\right) \, \partial_{\hat \rho} (\delta \phi) \right] \,,
\ea
where we use $\partial_\rho \phi = \partial_\rho (\delta \phi)$ to allow the use of the background metric. Similarly,
\ba
 \delta \left( \frac{2g_\ssR^2}{\kappa^2} \, e^\phi
 - \frac12\, \kappa^2 \cQ^2 \, e^\phi e^{-8W} \right) &=&
 \left( \frac{2g_\ssR^2}{\kappa^2} - \frac{\kappa^2}2 \, \cQ^2 \right)
 e^{\varphi_0} \delta\phi - e^{\varphi_0} \cQ^2 \kappa^2 \left(
 \frac{\delta\cQ}\cQ - 4 \delta W \right) \nn\\
 &=& - e^{\varphi_0} \cQ^2 \kappa^2 \left( \frac{\delta\cQ}\cQ - 4 \delta W \right) \,,
\ea
where the second line uses the rugby ball condition for the background value of $\cQ$. With these the dilaton equation becomes
\be
 \frac{ \partial_{\hat \rho} \left[ \sin \left( \hat\rho/ L \right)
 \partial_{\hat \rho} ( \delta \phi) \right]}{\sin \left( \hat\rho / L \right)}
 = - \cQ^2 \kappa^2 \left( \frac{\delta\cQ}\cQ - 4 \delta W \right)
 = - \frac1{L^2} \left( \frac{\delta\cQ}\cQ - 4 \delta W \right) \,,
\ee
which is the form used in the main text.

Similarly, the linearization of the Einstein equation, eq.~\pref{eomansatz}, uses
\ba
 \delta \left( \frac{g_\ssR^2 e^{\phi} }{\kappa^2} \right)
 &=& \frac14 \, \kappa^2 \cQ^2 e^{\varphi_0} \delta \phi
 = \frac{e^{\varphi_0}}{4L^2} \, \delta\phi \nn\\
 \delta \left( \kappa^2 \cQ^2 e^{\phi-8W} \right)
 &=& \kappa^2 \cQ^2 e^{\varphi_0} \left( \frac{2\delta\cQ}\cQ
 + \delta\phi - 8 \delta W \right)
 = \frac{e^{\varphi_0}}{L^2} \left( \frac{2\delta\cQ}\cQ + \delta\phi - 8 \delta W \right)\nn\\
 \delta (B')^2 &=& 2 \partial_\rho B_0 \, \partial_\rho (\delta B)
 = \frac{2 e^{\varphi_0}}{L^2} \, \cot \left( \frac{\hat\rho} L \right)
 \partial_{\hat \rho} (\delta B) \,.
\ea
Since $J$ is perturbatively small, the background metric can be used to simplify the current term in the action,
\be
 \delta( \sqrt{-g} \; J) = \delta( \sqrt{-\hat g} \; e^{4W +B} \, J)
 = \sqrt{-\hat g} \; e^{-\varphi_0} \alpha L \sin \left(
 \frac{\hat \rho} L \right) \, J \,,
\ee
ensuring that $J$ appears as a new contribution $\kappa^2 J$ to the Einstein equations, eq.~\pref{eomansatz}. Finally, the derivative terms for $\delta B$ become
\be
 \delta B'' + \frac2L \, \cot \left( \frac{\hat\rho} L\right) B'
 = e^{\varphi_0} \,
 \frac{ \partial_{\hat\rho} \left[\sin^2(\hat\rho/L) \partial_{\hat \rho}
 (\delta B) \right]}{\sin^2(\hat\rho/L)} \,.
\ee
Putting this all together yields
\ba
 \frac{ \partial_{\hat\rho} \left[ \sin^2 \left(\hat \rho / L \right)
 \partial_{\hat\rho} (\delta B) \right]}{\sin^2 \left(\hat \rho / L \right)}
 &=& - \frac1{L^2} \left[ \delta\phi + \frac32 \left( \frac{\delta\cQ}\cQ \right)
 - 6 \delta W + \kappa^2 J L^2 e^{-\varphi_0} \right]
 - \frac4L \, \cot \left( \frac{\hat\rho} L \right) \partial_{\hat\rho} W \nn\\
 \frac{ \partial_{\hat\rho} \left[ \sin^2 \left(\hat \rho /L \right)
 \partial_{\hat\rho} (\delta B) \right]}{ \sin^2 \left( \hat\rho /L\right)}
 &=& - \frac1{L^2} \left[ \delta\phi + \frac32\left( \frac{\delta\cQ}\cQ \right)
 - 6\delta W + \kappa^2JL^2 e^{-\varphi_0} \right] - \partial_{\hat\rho}^2 W \,,
\ea
and
\be
 \hat R = -4 e^{\varphi_0} \left[ \frac{2W}{L^2} + \frac1L \, \cot \left(
 \frac{\hat\rho} L \right) \partial_{\hat\rho} W
 + \partial_{\hat\rho}^2 W \right] + \frac{2e^{\varphi_0}}{L^2} \left(
 \frac{\delta\cQ}\cQ \right) - 2\kappa^2 J  \,,
\ee
which are the equations solved in the main body.

\section{Some useful integrals}

This appendix evaluates the integrals $\cM_i$ and $\cH_i$ encountered in the main text.

\subsection*{Evaluation of $\overline\cM$}
\label{app:findM}

This section evaluates the constant encountered in the $\varphi_1$ perturbation. The integrals of interest are
\ba
 \cM_1(x) &=& \int_0^x \exd y \; \sin^2 y \, \ln \abs{\frac{1-\cos y}{\sin y}} \nn\\
 \cM_2(x) &=& \int_0^x \exd y \; \frac{\cM_1(y)}{\sin^2 y} \nn\\
 \hbox{and} \quad
 \overline\cM &=& \int_0^\pi \exd x \; \sin x \, \cM_2(x) \,.
\ea
First of all, notice that the logarithm in the first line is antisymmetric under $y \rightarrow \pi-y$, while $\sin^2 y$ is symmetric. This means that the first integral integrates to $0$ if $x = \pi$: that is, $\cM_1(\pi) = 0$. These observations justify the following manipulations:
\ba
 \cM_1(\pi-x) &=& \cM_1(\pi) - \int_{\pi-x}^\pi \exd z \; \sin^2 z
 \ln \abs{\frac{1-\cos z}{\sin z}} \nn\\
 &=& - \int_{0}^x \exd y \; \sin^2(\pi-y) \,
 \ln\abs{ \frac{1-\cos(\pi-y)}{\sin(\pi-y)}} \nn\\
 &=& \int_0^x \exd y \; \sin^2 y \, \ln\abs{ \frac{1-\cos y}{\sin y}} \nn\\
 &=& \cM_1(x) \,.
\ea
The same manipulations applied to $\cM_2$ then give:
\ba
 \cM_2(\pi-x) &=& \cM_2(\pi) - \int_{\pi-x}^\pi \exd z \;
 \frac{\cM_1(z)}{\sin^2 z} \nn\\
 &=& \cM_2(\pi) - \int_{0}^x \exd y \;
 \frac{\cM_1(\pi-y)}{\sin^2(\pi-y)}\nn\\
 &=& \cM_2(\pi) - \int_{0}^x \exd y \;
 \frac{\cM_1(y)}{\sin^2 y} \nn\\
 &=& \cM_2(\pi) - \cM_2(x) \,.
\ea

Numerical evaluation of $\cM_2(\pi)$ is complicated by the weak convergence of the integral near $\pi$. It can be evaluated more efficiently by using the above expressions to relate it to $\cM_2(\pi/2)$. That is, numerical integration gives $\cM_2(\pi/2) = -0.5$ to within the numerical ({Maple 11}) precision. Using this, we find
\be
 \cM_2(\pi) = 2 \cM_2(\pi/2) = -1 \,.
\ee
Hence $\cM_2$ satisfies
\be
 \cM_2(\pi-x) = -1 - \cM_2(x) \,.
\ee

To evaluate $\overline\cM$, use
\ba
 \overline \cM &=& \int_0^{\pi/2} \exd x \; \sin x \, \cM_2(x)
 + \int_{\pi/2}^\pi \exd x \; \sin x \, \cM_2(x) \nn\\
 &=& \int_0^{\pi/2} \exd x \; \sin x \, \cM_2(x)
 + \int_{0}^{\pi/2} \exd x \; \sin(\pi-x) \, \cM_2(\pi-x) \nn\\
 &=& \int_0^{\pi/2} \exd x \; \sin x \, \cM_2(x)
 + \int_{0}^{\pi/2} \exd x \; \sin x \Bigl[ -1 - \cM_2(x) \Bigr] \nn\\
 &=& - \int_0^{\pi/2} \exd x \; \sin x
 = -1 \,.
\ea

\subsection*{Evaluation of $\overline \cH$}

Recall the definitions,
\ba
 \cH_1(x) &=& \int_0^x \exd y \; \sin^2 y \, \ln\abs{\sin y} \nn\\
 \cH_2(x) &=& \int_0^x \exd y \; \frac{\cH_1(y)}{\sin^2 y} \nn\\
 \bar\cH &=& \int_0^\pi \exd y \; \sin y \, \cH_2(y) \,.
\ea
In this case numerical evaluation gives (Maple 11):
\be
 \cH_1(\pi) = \frac\pi4 \, ( 1 - \ln 4) \,.
\ee
Similar numerical integration to evaluate $\cH_2(\phi)$ is complicated by the apparent singularity at the endpoints caused by the factors of $1/\sin^2 y$ in the integrand. These can be dealt with by repeating the arguments of the previous section, which in this case give
\ba
 \cH_1(\pi-x) &=& \cH_1(\pi) - \int_{\pi-x}^\pi \exd y \;
 \sin^2 y \, \ln\abs{\sin y} \nn\\
 &=& \cH_1(\pi) - \int_0^x \exd y \; \sin^2 y \, \ln\abs{\sin y} \nn\\
 &=& \cH_1(\pi) - \cH_1(x) \,.
\ea

Next consider the following symmetry properties of $\cH_2$:
\ba
 \cH_2 \( \frac\pi2 + x \) &=& \cH_2 \( \frac\pi2 \)
 + \int_{\pi/2}^{\pi/2+x} \exd z \; \frac{\cH_1(z)}{\sin^2 z} \nn\\
 &=& \cH_2 \( \frac\pi2 \) - \int_{\pi/2}^{\pi/2-x} \exd y \;
 \frac{\cH_1(\pi-y)}{\sin^2(\pi-y)} \nn\\
 &=& \cH_2 \( \frac\pi2 \) + \int_{\pi/2-x}^{\pi/2} \exd y \;
 \frac{\cH_1(\pi) - \cH_1(y)}{\sin^2 y} \,,
\ea
and simplify using
\ba
 \int_{\pi/2-x}^{\pi/2} \exd y \; \frac{\cH_1(y)}{\sin^2 y }
 &=& \int_{0}^{\pi/2} \exd y \; \frac{\cH_1(y)}{\sin^2 y}
 -\int_0^{\pi/2-x} \exd y \; \frac{\cH_1(y)}{\sin^2 y } \nn\\
 &=& \cH_2 \( \frac\pi2 \) - \cH_2 \( \frac\pi2-x \) \,,
\ea
to get
\ba
 \cH_2 \( \frac\pi2+x \) &=& \cH_2 \( \frac\pi2-x \)
 + \cH_1(\pi) \, \int_{\pi/2-x}^{\pi/2} \frac{ \exd y }{\sin^2 y} \nn\\
 &=& \cH_2 \( \frac\pi2-x \) + \cH_1(\pi) \, \cot \( \frac \pi 2 - x \) \,.
\ea

The evaluation of $\overline\cH$ now proceeds, with
\ba
 \overline \cH &=& \int_0^\pi \exd x \; \sin x \, \cH_2(x) \nn\\
 &=& \int_0^{\pi/2} \exd x \; \left\{ \sin x \, \cH_2(x)
 + \sin \( \frac\pi2 + x \) \, \cH_2 \( \frac\pi2+x \) \right\} \nn\\
 &=& \int_0^{\pi/2} \exd x \; \left\{ \sin x \, \cH_2(x)
 + \sin \( \frac\pi2-x \) \left[ \cH_2 \( \frac\pi2-x \)
 + \cH_1(\pi) \, \cot \( \frac\pi2-x \) \right] \right\} \nn\\
 &=& 2 \int_0^{\pi/2} \exd x \; \sin x \, \cH_2(x)
 + \cH_1(\pi) \int_0^{\pi/2} \exd x \; \cos \( \frac\pi2-x \) \nn\\
 &=& 2 \int_0^{\pi/2} \exd x \; \sin x \, \cH_2(x) + \cH_1(\pi) \,.
\ea
This can now be integrated numerically without problems, giving (to ten decimal places) a result consistent with $\overline \cH = -2 + \ln 4$.

\end{document}